\documentclass[pdflatex,sn-mathphys-num]{sn-jnl}

\usepackage{graphicx}%
\usepackage{multirow}%
\usepackage{amsmath,amssymb,amsfonts}%
\usepackage{amsthm}%
\usepackage{mathrsfs}%
\usepackage[title]{appendix}%
\usepackage{xcolor}%
\usepackage{textcomp}%
\usepackage{manyfoot}%
\usepackage{booktabs}%
\usepackage{algorithm}%
\usepackage{algorithmicx}%
\usepackage{algpseudocode}%
\usepackage{listings}%
\usepackage[section]{placeins}

\newcommand{\beq}{\begin{equation}}
\newcommand{\eeq}{\end{equation}}
\newcommand{\beqa}{\begin{eqnarray}}
\newcommand{\eeqa}{\end{eqnarray}}
\newcommand{\bd}[1]{ \mbox{\boldmath $#1$}}

\theoremstyle{thmstyleone}%
\theoremstyle{thmstyletwo}%

\theoremstyle{thmstylethree}%

\begin{document}
\def\ii{\'\i}

\title[The $0^+$-spectrum in rare earth nuclei within the pseudo-$SU(3)$ shell model]
{The $0^+$-spectrum in rare earth nuclei within the pseudo-$SU(3)$ shell model}


\author*[1,2]{\fnm{Peter O.} \sur{Hess}}\email{hess@nucleares.unam.mx}

\author[2,3]{\fnm{Sahila} \sur{Chopra}}\email{chopra.sahila@gmail.com}
\equalcont{These authors contributed equally to this work.}

\affil*[1]{\orgdiv{Institito de Ciencas Nucleares}, \orgname{Universidad Nacional Aut\'onoma de M\'exico}, \orgaddress{\street{Circito Exterior S/N}, \city{Mexico-City}, \postcode{04510}, \state{CDMX}, \country{Mexico}}}

\affil[2]{\orgdiv{Frankfurt Institute for Advanced Studies}, \orgname{J.W. von Goethe Univeristy}, \orgaddress{\street{Ruth-Moufang-Straße 1}, \city{Frankfurt am Main}, \postcode{60438}, \state{Hessen}, \country{Germany}}}

\affil[3]{\orgdiv{Maharishi Markandeshwar}, \orgname{Deemed to be University}, \city{Ullana-Ambala}, \postcode{133-207}, \country{India}}


\abstract{
The study of the structure of the $0^+$ spectrum in heavy nuclei has drawn much attention 
in the last two decades.
In this contribution we study their properties from a microscopic
point of view.
The pseudo-$SU(3)$ model ($\widetilde{SU}(3))$ is applied to some rare earth nuclei, namely to
Sm, Gd, Dy, Er, Yb and Hf isotopes. It is shown that the $0^+$ spectrum,
and the accumulation of states at certain energies, can be well 
understood using this microscopic model, which takes into account the {\it Pauli
Exclusion Principle} (PES). Intentionally, a very simple model Hamiltonian is applied and only
the valence shell is taken into account, in order to high-lighten certain cross
features. It is demonstrated that the microscopic Hilbert space is essential in
understanding the accumulation of $0^+$-states.
A discussion to other models is provided.
Also the dominance of $B(E2)$-transitions from the $\gamma$-band over those 
from the $\beta$-band turns out to be trivial, in contrast to within some collective models.
}





\maketitle

\section{Introduction}
\label{sec1}

In the recent couple of decades the interest in the spectrum and the  structure 
of $0^+$states has increased significantly,
as can be seen in a recent article \cite{ani2025},
and references therein, which gives an excellent summary of this
effort and provides a wealth of data for $0^+$ states in well deformed rare earth nuclei. 
A particular interest is in unraveling the nature of the $0^+$ states and how to
interpret them within collective models of nuclei, as 
the geometric collective model of nuclei
\cite{Eisenberg,GCM-Frankfurt} or the {\it Interacting Boson Approximation} (IBA)
\cite{IBA-I}. In the geometric model, the low lying $0^+$ states are interpreted as
$\beta$-vibrations, oscillatory motions along the symmetry axis, while the asymmetric oscillations are  related to $\gamma$-vibrations. After
a geometric mapping, the same interpretation is given in
the IBA model, where here we restrict to the first version of it, mostly used in literature. Both models are able to 
describe different deformations of nuclei, spherical or deformed. However, they are 
different in the context of their formulation. The IBA is an algebraic model 
with nucleon pairs as its basic degrees of freedom,
while the geometric model describes surface vibrations with no microscopic relation.
Nevertheless, it was shown multiple times that both models are intimately related mathematically
(see as examples \cite{mosh1980,symmetry}). In fact, as showm in \cite{symmetry}
 mathematically these models are equal, their difference is in the choice of the Hilbert space. 

A particular property for well deformed nuclei is the difference in intensity of the $B(E2; 0_\beta^+ \rightarrow 2_g^+)$
and $B(E2;2_\gamma \rightarrow 0_g^+)$, here the index $g$ refers to the gound state band,
$\gamma$ to the gamma band and $\beta$ to the $\beta$-band. The intensity of the first one
is usually smaller than  the second one, which is reverse in the geometric model. 
In addition, within the deformed limit
($SU(3)$) of the IBA there is no
transition between the $\gamma$-band and the ground state band, while there is one
within the geometric model for large deformed nuclei. In order to adjust the 
IBA to experiment,
one has to use a symmetry breaking Hamiltonian \cite{ani2025}.

The reason for the last observation is discussed in \cite{symmetry}: Within the rotational limit n($SU(3)$) of the IBA the lowest band is given by the 
{\it irreducible representation} (irrep) $(2N,0)$, $N$ being the number of bosons. Thus,
the $\gamma$-band belongs to a different $SU(3)$ irrep and no quadrupole transition
is possible. However, the main problem was demonstrated by K.T. Hecht 
\cite{Hecht-IBA}, namely
that the IBA ignores the effect of the PES, treating the bosons as
structureless and not taking into
account their substructure, which gets important in the middle of a shell. 
If the substructure is taken into account, the lowest irrep is
of the type $(\lambda , \mu )$ with both numbers different from zero and even.  
With the corrected Hilbert space,
the $\gamma$-band is now part of the lowest irrep and thus transitions are possible,
while the $\beta$-band belongs to a different irrep and transitions are not possible.
Deviating a bit form the $SU(3)$ limit, then permits small transitions from the
$\beta$-band to the ground state band, but not dominating the transitions from the 
$\gamma$-band. Thus, such a simple observation resolves the problem 
encountered in the IBA. 

This problem was also exposed in \cite{symmetry} and
called in \cite{draayer-book} as the {\it fatal flaw} of the IBA. As shown in
\cite{ani2025}, applying the IBA the transitions can be described by symmetry
mixing within the IBA. This can be considered a patch to get a better agreement. 

But even when both collective models can describe many effects of the collective spectrum,
they do not reproduce the dense accumulation of $0^+$ states at
relatively low energy. The motivation for this contribution is to understand the origin and the
{\it basic} structure of these states. 
For this purpose, it is often very useful not to consider an involved, 
complete Hamiltonian, but rather simplifies one which already reflects the main 
properties of the system.
We apply the 
$\widetilde{SU}(3)$ model in order to show that a microscopic model, based on the
microscopic space of the shell model, is needed to describe the complicated 
$0^+$ séctrum. Our intention is not to provide an involved theory with a complete
realistic interaction, which only would obscure the broad picture. 
Rather, we are interested in the origin of the $0^+$ spectrum, i.e., 
providing an underpinning of the observed spectrum.
In fact, the microscopic Hilbert space plays a more dominant role in 
describing the accumulation of $0^+$-states.

The present investigation can also be done using the proxy-$SU(3)$ model
\cite{proxy}, which includes part pf the intruder levels..

The presentation is as follows: In section \ref{sec2} the $\widetilde{SU}(3)$ model 
\cite{hecht,arima} is  presented. In section \ref{sec3} the simplified model Hamiltonian
is given and an explanation is provided on how the spectrum of a particular nucleus
is obtained and adjusted, In section \ref{sec4} the results are presented and discussed.
Finally, in section \ref{sec5} Conclusions are drawn.

\section{The $\widetilde{SU}(3)$ model}
\label{sec2}

The approach to $\widetilde{SU}(3)$ is as follows:
In each harmonic oscillator shell $\eta$
the orbital belonging to the
largest spin $j=\eta + \frac{1}{2}$ is removed from
consideration as an active orbital, it is considered
as a spectator.
To the remaining orbitals the redefinition

\beqa
j ~=~ l\pm\frac{1}{2} & \rightarrow & {\tilde l}~=~ 
l\mp\frac{1}{2}
~,~
\eta ~ \rightarrow ~ {\tilde \eta} ~=~ \eta - 1
~~~,
\label{eq-1}
\eeqa
is applied, where ${\tilde l}$ denotes the
{\it pseudo-orbital angular momentum}
and ${\tilde \eta}$ the pseudo-shell number. With
this redefinition alone it is easily verified that
each shell $\widetilde{\eta}$ has the same content
as the corresponding shell in the standard $SU(3)$ 
model.

For large deformations, the Nilsson states for axial symmetric 
nuclei are classified by their asymptotic quantum numbers
\cite{ring}

\beqa
\Omega \left[ \eta\eta_z\Lambda \right]
~~~,
\label{nil-1}
\eeqa
where $\eta_z$ is the oscillation number 
in the $z$-direction, $\Lambda$ is the projection 
of the orbital
angular momentum onto the same axis and $\Omega$
= $\Lambda \pm \frac{1}{2}$.  

Excluding the intruder levels, the 
reassignment of the orbitals is

\beqa
\Omega ~=~ \Lambda \pm\frac{1}{2} & \rightarrow & 
\widetilde{\Lambda}~=~ 
\Lambda\mp\frac{1}{2}
~~~.
\label{eq-1a}
\eeqa
Inspecting the Nilsson diagrams \cite{ring},
those orbitals with {\it the same quantum numbers}
$\left[\widetilde{\eta}\widetilde{\eta_z}
\widetilde{\Lambda}\right]$ are degenerate,
which implies a very small pseudo-spin-orbit interaction
and as a consequence an {\it approximate symmetry}. In addition, the content
of the $\widetilde{\eta}$ shell corresponds to the 
same one in the
standard shell model.
Thus, the shell model 
can be directly extended to heavy nuclei, 
using the $\widetilde{SU}(3)$ model. 

The basis in the Hilbert space is a direct product
of the $\widetilde{SU}(3)$ states and the ones from
the unique (intruder) orbitals. 
Their contribution to the
nuclear dynamics is taken into account by 
a well defined scaling factor (effective charge).

That the nuclear force exhibits
such an unexpected symmetry is today well understood,
parting from microscopic field theoretic models of 
the nuclear interaction \cite{fieldsu3} 
and mapping to the effective nuclear
interaction. The complete Hilbert space of the shell model 
is a direct 
product of a state described within a $\widetilde{SU}(3)$,
in the same manner as the $SU(3)$ model of Elliott
for light nuclei, and a state describing the nucleons in
the intruder orbitals. The nucleons in the intruder orbitals 
are assumed to play the minor role of an observer: Nucleons
in the unique orbitals play a passive role
in the dynamics due to their opposite parity and their
coupling to spin-zero pairs, 
contributing only to the binding energy. The pairing energy
of nucleons in the unique orbitals is big, due
to the large $j$ \cite{ring}, compared to 
the nucleons in the normal
orbitals. Thus, nucleons in the unique orbitals only
contribute at high energies, e.g., in the back-bending 
effect and related phenomena \cite{draayer-book}.  
The contribution of the nucleons in the unique orbitals
are treated via well defined effective charges for electromagnetic transitions 
\cite{NPA1994}. 
In conclusion, the restriction to the $\widetilde{SU}(3)$
is well justified for states at low energy.
Though, the $\widetilde{SU}(3)$ has its limits, as not including independent
dynamical effects of nucleons in the unique orbitals, it is still useful,
as we will see further below.

The effectiveness of $\widetilde{SU}(3)$ was
demonstrated in \cite{NPA1994}, for the case of
the pseudo-symplectic model of the nucleus
\cite{pseudo-sympl} and many other applications of the 
$\widetilde{SU}(3)$ model (see, for example, 
\cite{cseh-quart}).

Instead of the $\widetilde{SU}(3)$ the proxy-$SU(3)$ model \cite{proxy} can be 
applied, too, which takes in addition into account part of the intruder levels.
It would be interesting to use this model, because it involved a more
extended, dynamical model space.

\subsection{Construction of the model space}
\label{model-space}

Here, we expose the path on how to deduce the 
$\widetilde{SU}(3)$ shell model space: Each shell
$\eta$ has $\frac{1}{2}(\eta +1)(\eta +2)$
orbital degrees of freedom. Taking into account the two
spin degrees of freedom, the group-chain classifying
the states within the shell $\eta$ (either for protons or neutrons) is given by

\beqa
U((\eta +1 )(\eta + 2)) & \supset~~~
U(\frac{1}{2}(\eta + 1)(\eta + 2)) \otimes U_S(2)
\nonumber \\
\left[ 1^N\right] & ~~~~~~~~~~~ \widetilde{\left[ h\right]}
~~~~~~~~~~~~~~~~~~~ \left[ h\right]
\nonumber \\
U(\frac{1}{2}(\eta + 1)(\eta + 2)) & \supset ~~~SU(3)
\nonumber \\
\left[ h\right] & ~~~~~~~~~~~ (\lambda , \mu )
~~~,
\label{class}
\eeqa
where $[h]$ is a short hand notation for 
$\left[h_1,h_2\right]$
and the tilde denotes the conjugate Young diagram where
rows and columns are interchanged. (\ref{class})
gives the relation of the spin-part (denoted by the index $S$)
and the orbital part, such that the complete state is
anti-symmetric ($\left[ 1^N\right]$, with $N$ as the
number nucleons in the shell $\eta$). 
In heavy nuclei the protons and neutron
have to be considered within different shells, which is the
reason why (\ref{class}) is used.
For the reduction to
$SU(3)$ programs are available
\cite{bahri}.

When dealing with heavy nuclei one encounters 
an exploding number of states ($SU(3)$ irreps).
Thus, one not only has to find a consistent path
on how to determine the lowest states, one also has to
introduce a cut-off:

\begin{itemize}

\item Because protons and neutrons are in different shells, the 
construction of the model space is applied separately to them,
as done {\it in any shell model application for heavy nuclei}
\cite{ring}. The protons and neutrons move in the
same mean field.

\item In a final step, the proton and neutron part are 
combined, proposing a particular selection, whose
origin is the  demand that the proton and neutron fluid
are aligned. In \cite{draayer1} this is discussed
and in more detail in \cite{NPA576}: In the nuclear
shell model the aligned irreps
correspond to
aligned principle axes of the two (proton-neutron) rotors.
All other irreps are non-aligned proton-neutron
rotors, which lie at higher energy, corresponding to
scissors mode, or isovector resonances \cite{Eisenberg}.
Thus, the restriction to aligned proton-neutron irreps
is a pretty good approximation, taking effectively into
account the interaction, which
tends to align the proton and neutron fluid.

I.e., a linear coupling between the proton and neutron fluid is 
applied, namely

\beqa
(\lambda_p,\mu_p) \otimes (\lambda_n,\mu_n)
& \rightarrow &
(\lambda_p + \lambda_n,\mu_p + \mu_n)
~~~.
\label{linear}
\eeqa

\end{itemize}

\section{The model Hamiltonian and adjusting spectra}
\label{sec3}

As the model Hamiltonian we use a particular simple one, only depending of
Casimir operators of $\widetilde{SU}(3)$, no symmetry
breaking allowed. Also, the calculations take into account only the valence shell,
i.e., there will be no $\hbar\omega$ term.

There is another property of this simplified model: $0^+$-states with the
{\it same} $(\tilde{\lambda},\tilde{\mu})$ are at the same energy. i.e., when an
irrep has the multiplicity $k$, there are $k$ degenerate states at the same energy,
for example $k$ $0^+$ states, when ${\tilde \lambda}$ and $\tilde{\mu}$ are both even.
That is, there is a $k$-fold degeneracy of $0^+$-states.
The reduction is taken from \cite{elliott}.

The model Hamiltonian is

\beqa
{\bd H} & = & \chi {\bd C}_2(\widetilde{SU}(3)) + (a+a_L (-1)^L) {\bd L}^2
+ b {\bd K}^2 + c \bd{C}_3(\widetilde{SU}(3)) + d \left( \bd{C}_2(\widetilde{SU}(3))\right)^2
~~~,
\label{ham}
\eeqa
where ${\bd C}_k(\widetilde{SU}/3))$ ($k=2,3$) is the $k$'th order Casimir operator of 
$\widetilde{SU}(3)$.

The eigenvalues of the operators are

\beqa
{\bd L}^2 & \rightarrow & L(L+1)
\nonumber \\
{\bd K} & \rightarrow & K
\nonumber \\
{\bd C}_2 (\widetilde{SU}(3)) & \rightarrow & \tilde{\lambda}^2 + \tilde{\lambda}\tilde{\mu} 
+ \tilde{\mu}^2
+3\tilde{\lambda} + 3\tilde{\mu}
\nonumber \\
{\bd C}_2 (\widetilde{SU}(3)) & \rightarrow & (\tilde{\lambda} - \tilde{\mu})
(2\tilde{\lambda} + \tilde{\mu} + 3)
(\tilde{\lambda} + 2\tilde{\mu} + 3 )
~~~.
\label{eigenvalues}
\eeqa
The $(\tilde{\lambda},\tilde{\mu})$ are the quantum labels of the 
$\widetilde{SU}(3)$ group, $L$ is the angular momentum and $K$ is the projection of
the angular momentum onto the intrinsic $z$-axis, where the 
operator is defined in \cite{pseudo-sympl}. The ${\bd K}^2$
operator lifts the degeneracy of angular momentum states with the same $L$, within
the same irrep.

For the algebraic quadrupole operator we use the definition of \cite{escher}:

\beqa
{\bd Q}^a_{2m} & = & \sqrt{3} {\bd C}^{(1,1)}_{2m}
~~~,
\label{Q-op}
\eeqa
where the index "a" refers to the algebraic part of 
the quadrupole operator, only
acting within the valence shell and ${\bd C}^{(1,1)}_{2m}$
is an $\widetilde{SU}(3)$ generator.

Its reduced matrix elements are given by

\beqa
& \langle (\tilde{\lambda}^\prime , \tilde{\mu}^\prime ), \kappa^\prime L^\prime \mid\mid
{\bf C}^{(1,1)}
\mid\mid \tilde{\lambda} , \tilde{\mu} ), \kappa L \rangle ~=~
\nonumber \\
& \sum_\rho\langle {\lambda}^\prime , \tilde{\mu}^\prime ), \kappa^\prime L^\prime \mid\mid
\tilde{\lambda} , \tilde{\mu} ), \kappa L \rangle_\rho
\langle  (\tilde{\lambda}^\prime , \tilde{\mu}^\prime )
\mid\mid\mid {\bd C}^{(1,1)} \mid\mid\mid
(\tilde{\lambda} , \tilde{\mu} ) \rangle_\rho
~~~,
\label{Q-ME}
\eeqa
where the first factor within the sum is the $SU(3)$-isoscalar factor 
\cite{bahri,escher} 
and the matrix element with three vertical lines on
each side refers to the triple reduced matrix element \cite{bahri,escher}.
The multiplicity label $\rho$ is 1, because the quadrupole operator is a generator of the group and only
connects to the same irrep.

The triple reduced element is given by \cite{escher}

\beqa
\langle  (\tilde{\lambda}^\prime , \tilde{\mu}^\prime )
\mid\mid\mid {\bd C}^{(1,1)} \mid\mid\mid
(\tilde{\lambda} , \tilde{\mu} ) \rangle
& = & (-1)^\Phi \sqrt{2C_2((\tilde{\lambda}^\prime , \tilde{\mu}^\prime ))}
\delta_{(\tilde{\lambda}^\prime , \tilde{\mu}^\prime ),(\tilde{\lambda} , \tilde{\mu} )}
\nonumber \\
\Phi & = & \left\{
\begin{array}{c}
1, ~{\rm for}~\tilde{\mu} \neq 0 \\
0, ~{\rm for}~\tilde{\mu} = 0
\end{array}
\right.
\label{trip-Q}
\eeqa

For the calculation of the $B(E2)$-values, the algebraic quadrupole operator
is multiplied by an effectove charge $e_{\rm eff}$, which is adjusted to the transition
$2^+_g \rightarrow 0^+_g$.
The parameter values, obtained through a fit, are listed in the Appendix 
\ref{app-Parameters},
Table \ref{e-eff}.. 

\FloatBarrier
\subsection{Obtaining occupation numbers}
\label{sub-ocu}

\begin{figure}
\centering
\includegraphics[width=1.\textwidth]{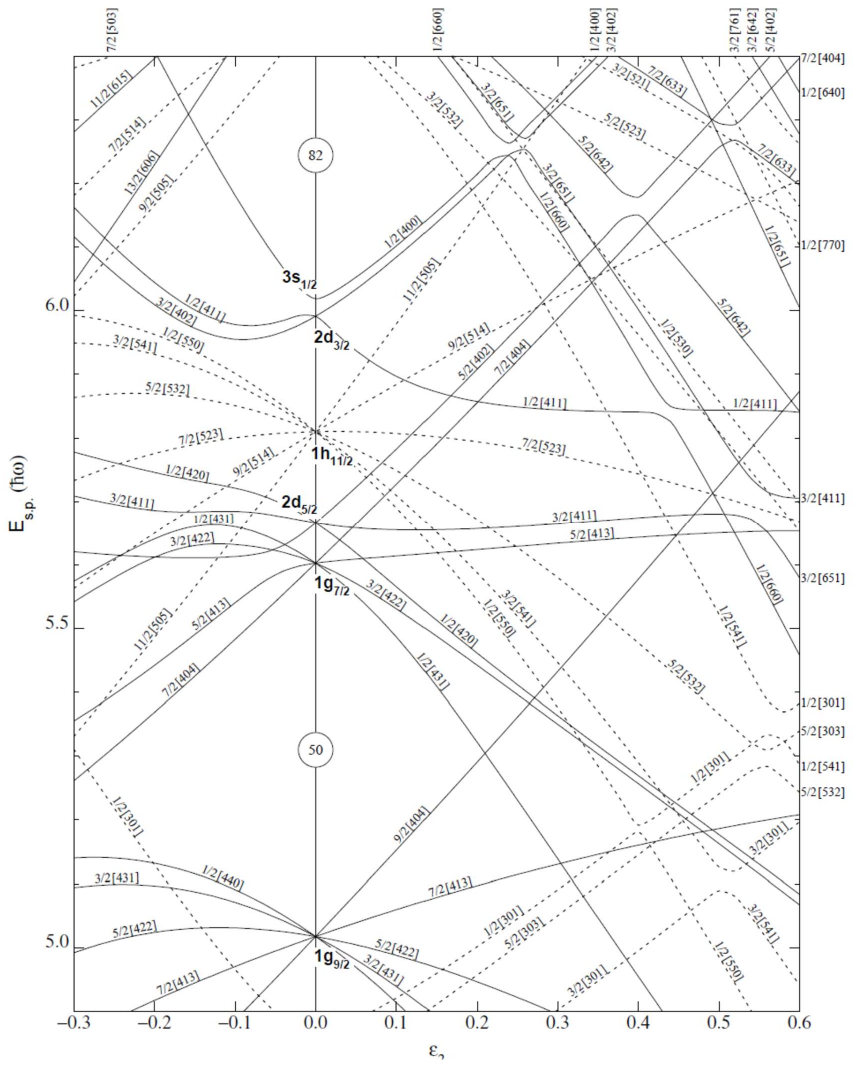} 
\caption{
Nilsson diagram for protons between the closed shells 50 and 82.
The figure is taken from \cite{nix-tables}.
} 
\label{Nil-p}
\end{figure}

\begin{figure}
\centering
\includegraphics[width=1.\textwidth]{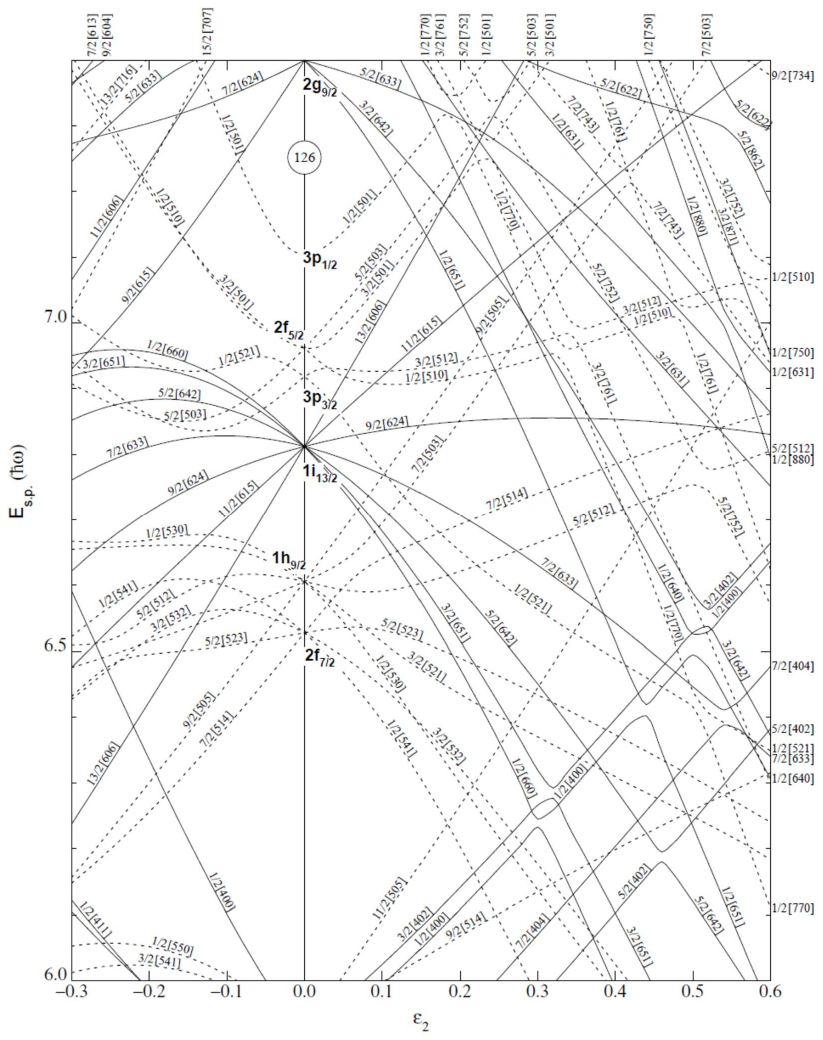} 
\caption{
Nilsson diagram for neutrons between the closed shells 82 and 126.
The figure is taken from \cite{nix-tables}.
} 
\label{Nil-n}
\end{figure}

The numbers of nucleons in the normal and unique orbitals are obtained as follows:

The nucleons are filled in within the Nilsson diagrams
at a given deformation value $\epsilon_2$, listed in \cite{nix-tables}
and in Appendix A. These values have 
a well defined relation to the quadrupole deformation value $\beta_2$, aso listed in
\cite{nix-tables}. The protons and neutrons are in different shells.
The Nilsson diagrams used here are depicted in Fig. \ref{Nil-p} for protons
and Fig. \ref{Nil-n} for neutrons, all figures taken from \cite{Nilsson-dia}.
Each orbital is occupied twice (one spin up and
one spin-down nucleon). The sum of nucleons in the normal orbitals gives the valence normal number
of nucleons for each sector. Adding the number of nucleons in the lower pseudo-shells gives
the total number of nucleons in normal orbitals. the same for the nucleons in the
unique orbitals. The results are listed in the tables in Appendix \ref{app-Parameters},
Table \ref{tab-nor}.

The number of nucleons in the valence shell determines the $\widetilde{SU}(3)$ content, for
protons and neutrons separately, using the reduction scheme given in (\ref{class})
and the programs provided in \cite{bahri}. 
The result is coupled in linear form as described above. 

The list of irreps obtained, given in Appendix \ref{appQN}, is a short one, 
only listing the ones with the
largest irrep. The complete list is quite large, but can be obtained on request.

\FloatBarrier
\section{Application to Sm, 
Gd, Dy, Er, Yb and Hf isotopes}
\label{sec4}

When applied to the aforementioned nuclei, the parameters of the Hamiltonian are
adjusted to the spectra of these nuclei. The optimal parameters are listed in
the Appendix \ref{app-Parameters}Table \ref{para}. 

The adjustment takes into account
the position of the first $2^+_g$ ($2_1^+$) and the first $6^+_g$ ($6_1^+$) 
ground band states.
Also, the position of the first $2^+_\gamma$
and $4^+_\gamma$ ($4_2^+$) are adjusted. 

There might appear a problem of identification: The states are associated to
their bands following the $B(E2)$-transitions. Sometimes, when for example
two $2^+$-states are near in energy, one may belong to the irrep with the
ground state band and the other, the lower lying one, belongs to another irrep.
In such a case, the identification of the $2^+_\gamma$-state has to be repositioned.
Still, in the table it is denoted as $2^+_2$, though a note will be added in the
corresponding column, when the association of the $2^+_2$ and $2^+_3$ states
has to be switched. The same is done for the $4_\gamma$-state

The data were retrieved from the {\it National Nuclear Data Center} \cite{brook}.

With respect to the $0^+_k$ excited states, specially of the
first $\beta$-band, the energies are taken from \cite{ani2025}.

On the theoretical side, we note that states with the same 
$(\tilde{\lambda} , \tilde{\mu} )$
have the same energy. However, as can be seen from the tables in the Appendix
\ref{appQN}, where for each nucleus a restricted list of $\widetilde{SU}(3)$ irreps is given,
the index attached to $(\tilde{\lambda} , \tilde{\mu} )$ refers to the degeneracy of these
irreps, i.e., when the degeneracy is $k$ there are $k$ $0^+$ states, which
represent a $k$-fold degeneracy in the theoretical spectrum. This degeneracy should be
lifted by adding more realistic interactions. However, as noted in the introduction,
we want to keep the theoretical part as simple as possible. It will imply an
association of a group of $k$ $0^+$ states  in the experimental spectrum.
For example, when after the ground state ($k=1$) an irrep appears with degeneray 2, then the next two $0^+$ states are associated to the double appearance of the irrep. 
Also, if in case the degeneracy is 8, then the next eight $0^+$ states are associated 
to the same irrep with the 8-fold multiplicity. Accordingly, when
the theoretical spectrum is compared to the experiment \cite{ani2025},
the experimental energies (see Appendix \ref{appenergies}) 
are grouped, calculating their
average value, taking the data from \cite{ani2025}, and the corresponding result is denoted by a star.  In case the experimental energies
can only be associated to less than $k$ states, the average is built on the number of 
the remaining listed states 
\cite{ani2025}, resulting in a lower limit to the actual experimental energy average.
In the extreme situation, when only one state is available, the "average" number is just
this number listed. 

For the transitions, only the $B(E2; 2_g^+ \rightarrow 0_g^+)$ transition is 
adjusted. The absolute values obtained are in the range of experiment, but more 
importantly their relative behavior is well reproduced.The effective charges, obtained in this fit, are also listed in the Appendix 
\ref{app-Parameters},Table \ref{e-eff}. It is noteworthy that all effective charges are
near the value $\frac{1}{2}$. 

The fits obtained are listed in the tables of Appendix \ref{appenergies}. The
excited $0^+$-states are separately listed and compared to experiment. As can be seen,
the fits are fairly good, considering the simplicity of the model. Looking at the $0^+$ spectrum,
the model can quite well reproduce their position and, especially the large
grouping (degeneracy) of those. The fits get less good when the proton or neutron shell
is near a closed shell, as for example $^{146}_{62}$Sm$_{84}$ where the number of
neutrons is near the shell closure of 82.
Near to a shell closure, the deformation limit of a pure 
$\widetilde{SU}(3)$ is not a good
symmetry. 

\subsection{Detailed discussion of the results}
\label{detail}

For each isotopic chain a representative nucleus was chosen whose 
$0^+$ spectrum is plotted.
These representative nculei are $^{150}$Sm, $^{158}$Gd, $^{162}$Dy,
$^{168}$Er, $^{172}$Yb and $^{178}$Hf. The low lying spectrum with the first bands 
are listed in the tables in Appendix C and the $B(E2)$ transitions in 
Appendix \ref{appBE2}.
The agreement in general is quite good, except near the closed physical neutron shel 
in the Sm-isotopes.

Here, we will concentrate on the discussion of the $0^+$ spectrum.

\begin{figure}
\centering
\includegraphics[width=0.45\textwidth]{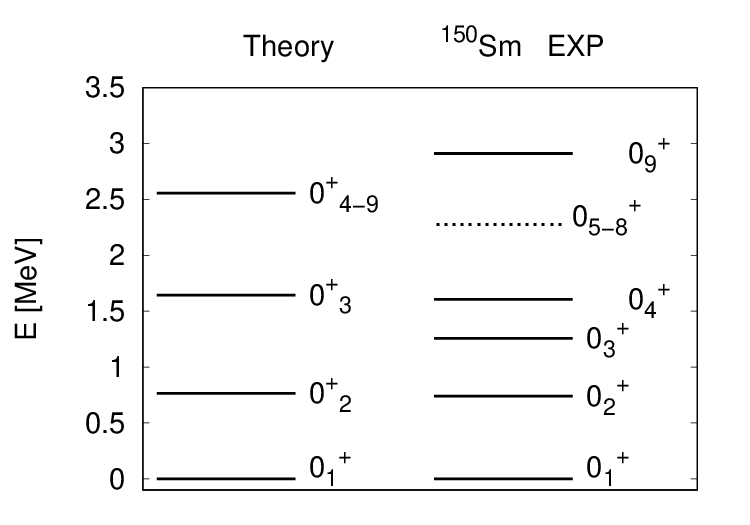} 
\caption{
Comparison of calculated and measured 0+ energy levels for $^{150}$Sm. Theoretical values are shown on the left; experimental results are on the right.
} 
\label{150-Sm}
\end{figure}

The $0^+$-spectrum of $^{150}$Sm is depicted in Fig. \ref{150-Sm}. We note a near degeneracy in experiment for the $0^+_{5-8}$ states, indicated by a dotted line.
The $0^+_4$ and $0^+_9$ states are below and above this nearly degenerated state,
respectively. Inspecting the tables for the irreps of $^{150}$Sm in 
Appendix \ref{appQN}, Table \ref{Sm-irrep},
there are the following irreps which contain $0^+$-states, ordered according to energy:
(28,4)$_1$, (30,0)$_1$,(20,10)$_1$, (24,6)$_9$. The first three are associated to the first three $0^+$ states in the spectrum. The (24,6) irrep is nine-fold degenerated,
i.e., the theory requires 9 near lying states in the experimental spectrum, 
which is indeed observed. This gives a clear hint that the microscopic shell model
can describe the degeneracy of the $0^+$ states in this nucleus and
that the degeneracy of a $(\tilde{\lambda},\tilde{\mu})$ irrep does account
for the accumulation of $0^+$-states observed.

Next, we look at the $^{158}$Gd nucleus, whose $0^+$ spectrum is depicted in
Fig. \ref{158-Gd}. Again, consulting the table in Appendix 
\ref{appQN}, Table \ref{Gd-irrep}, the irreps
containing $0^+$ states are (28,8)$_1$, (30,4)$_2$, (32,0)$_1$, (24,10)$_8$.
The (30,4) irrep is two-fold degenerate, which is well reproduced in the figure, associating the $0^+_2$ and $0^+_3$ in the experimental spectrum to this irrep. 
The $0^+_4$-state is associated to ((32,0), which appears only once. The 
(24,10) irrep is eight-fold degenerated, which explains the accumulation
of these state sin the experimental spectrum.

\begin{figure}
\centering
\includegraphics[width=0.45\textwidth]{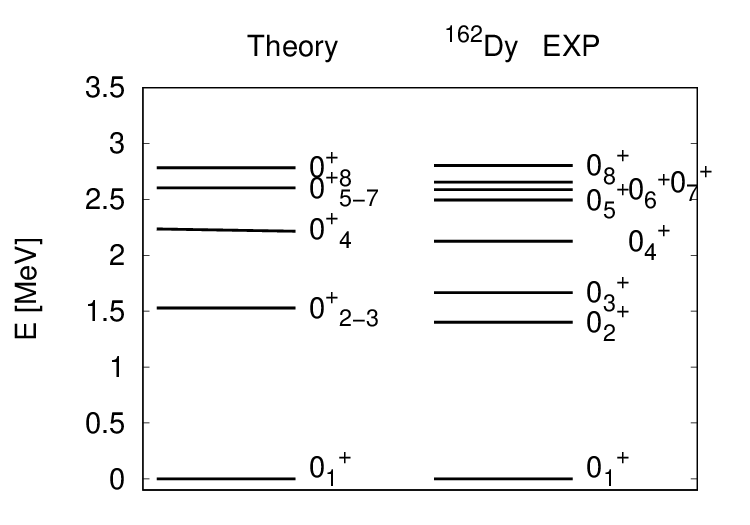} 
\caption{
Comparison of calculated and measured 0+ energy levels for $^{162}$Dy. Theoretical values are shown on the left; experimental results are on the right.
} 
\label{162-Dy}
\end{figure}

The $^{162}_{66}$Dy isotope was investigated in \cite{ani2025} within the IBA. Inspecting
the possible irreps within $\widetilde{SU}(3)$, using the tables in the 
Appendix \ref{appQN}, Table \ref{Dy-irrep},
we have  $(30,8)_1$, $(32,4)_2$, definitively grouping the ground state and $\gamma$-band 
into the $(30,8)$ irrep, while the $\beta$-band belongs to the $(32,4)$ irrep. 
The  $0^+$-spectrum for this nucleus is depicted in Fig. \ref{162-Dy}. The accumulation
of the $0^+$ states is reproduced quite well. This grouping is the result of the
$\widetilde{SU}(3)$-content and reflects the degeneracy quite well. It implies that the
$\widetilde{SU}(3)$ scheme is sufficient to explain these accumulation of states.
This example proofs that one does not need to break the 
$\widetilde{SU}(3)$ symmetry, in order to
reproduce the experimental data,
as the $B(E2)$ transition values in Appendix 
\ref{appBE2}. On he contrary, it shows a serious defect of the IBA,
which is forced to break the symmetry in order to explain the $B(E2)$-transitions.

\begin{figure}
\centering
\includegraphics[width=0.45\textwidth]{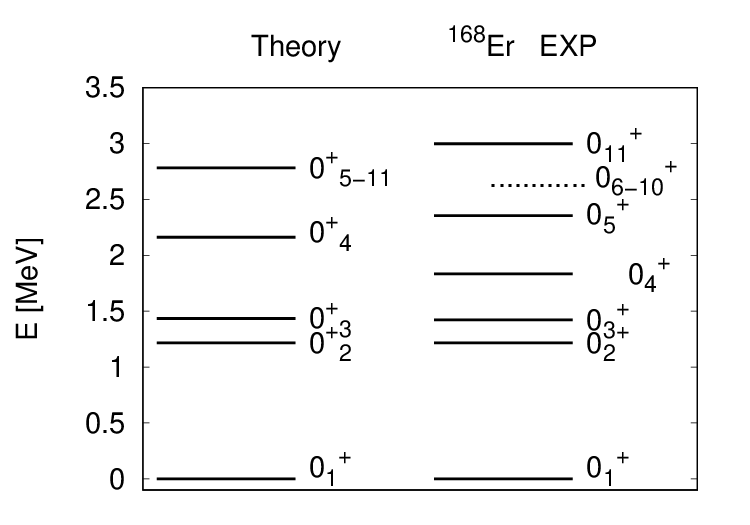} 
\caption{
Comparison of calculated and measured 0+ energy levels for $^{168}$Er. Theoretical values are shown on the left; experimental results are on the right.
} 
\label{168-Er}
\end{figure}

For $^{168}$Er we have to expand the list of irreps up to (28,10)$_8$, which is not listed in the tables. Only the irreps (34,4)$_1$, (36,0)$_1$, (26,14)$_1$ and
(20,20)$_1$ are listed, all non degenerate. This degenerate structure is also seen 
in the spectrum, see Fig. \ref{168-Er}: The first four $0^+$ states are well associated and at 
higher energy there is a clear reproduction of the accumulation of the next eight
$0^+$-states. 

\begin{figure}
\centering
\includegraphics[width=0.45\textwidth]{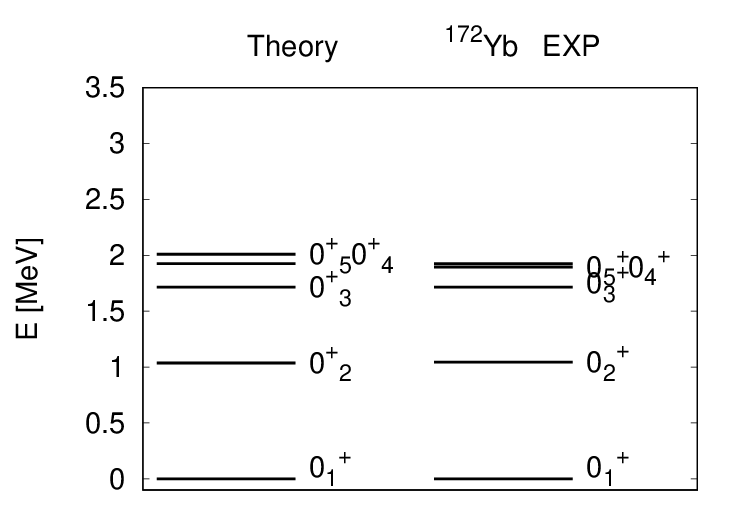} 
\caption{
Comparison of calculated and measured 0+ energy levels for $^{172}$Yb. Theoretical values are shown on the left; experimental results are on the right.
} 
\label{172-Yb}
\end{figure}

In $^{172}Yb$ the sequence of irreps, which contain $0^+$ states, 
is (12,28)$_1$, (32,6)$_1$, (34,2)$_1$, (24,16)$_1$ and (18,22)$_2$. Only the last 
irrep is doubly degenerate, which is well reflected in the spectrum
in Fig. \ref{172-Yb}. Otherwise $^{172}$Yb is quite uneventful.
It is also noted that in this isotope we see  a transition from a prolate 
($\tilde{\lambda} > \tilde{\mu}$) to oblate ($\tilde{\lambda} < \tilde{\mu}$)
deformation in the ground state.

\begin{figure}
\centering
\includegraphics[width=0.45\textwidth]{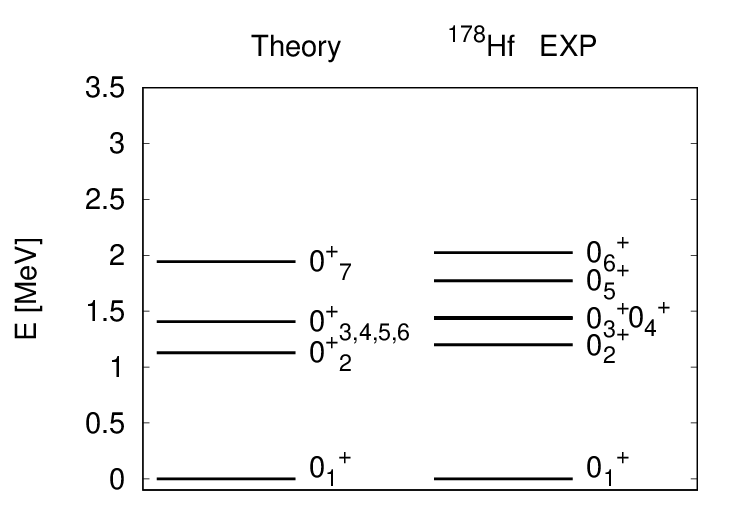} 
\caption{
Comparison of calculated and measured 0+ energy levels for $^{178}$Hf. Theoretical values are shown on the left; experimental results are on the right.
} 
\label{178-Hf}
\end{figure}

The last representative nucleus is $^{178}$Hf, Fig. \ref{178-Hf}. The sequence of 
irreps, containing $0^+$-states, is (6,32)$_1$, (2,34)$_1$, (12,26)$_4$ and 
(24,14)$_1$. First, the ground state and the next four excited irrep are 
already oblate, while the (24,14) irrep is still prolate, i.e,
we have a mixture of deformations in the same nucleus, hinting to local
minima in the geometric model at different deformations. In Fig. \ref{178-Hf}, while the first two $0^+$ state sare single degenerate, there is an accumulation of
$0^+$ states corresponding to the (12,26) irrep, though less  concentrated as in the other examples.

Consulting the tables in Appendix C for $0+$-states in other nuclei, we find the model accurately predicts them, including how many states showing the degeneracy. The rotational band pattern is also predicted well.

In conclusion, we can say that the accumulation of $0^+$ states at certain energies
has its origin in the degeneracy of $\widetilde{SU}(3)$ irrep of the microscopic
shell model. Also the transition from prolate to oblate states is predicted, 
suggesting different local minima in the geometric model.
We see reflected the  structure of degeneracies. The spectrum of all isotopes
were also reproduces quite well, not so clear near the neutron shell closure of 82.

The selected $B(E2)$ values, referring to the transition from the $\gamma$- to the
ground state band, are also well reproduced in their relative intensities. It is
interesting to note that the effective charges obtained, listed in 
Appendix \ref{app-Parameters}, are all
near $\frac{1}{2}$.

\begin{figure}
\centering
\includegraphics[width=0.45\textwidth]{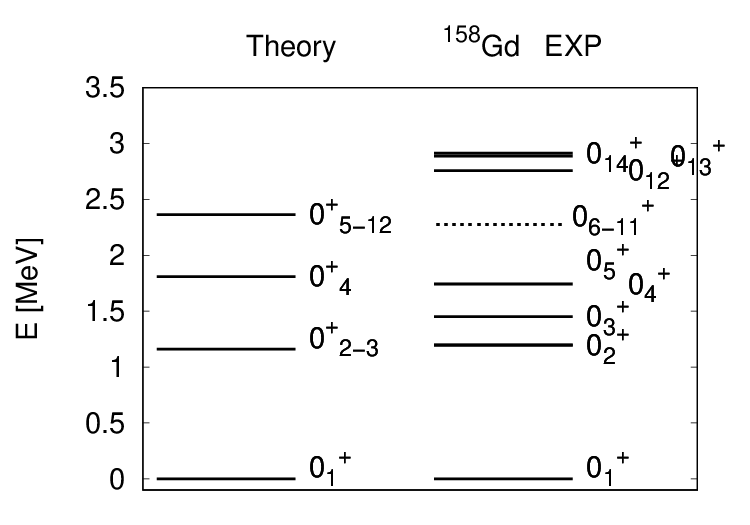} 
\caption{
Comparison of calculated and measured 0+ energy levels for $^{158}$Gd. Theoretical values are shown on the left; experimental results are on the right.
} 
\label{158-Gd}
\end{figure}

\FloatBarrier
\section{Conclusions}
\label{sec5}

The main motivation of this contribution was to demonstrate that a simple model
for heavy nuclei, base on the shell model, is able to describe the spectrum of
the $0^+$-states with a rather good quality. The $\widetilde{SU}(3)$ model was used and
the content of the irreps $(\tilde{\lambda},\tilde{\mu})$ was determined. 
In particular, these irreps may appear several times (degeneracy),
which results in an accumulation of $0^+$ states, just what is observed.
Also, the spectrum of low lying bands is well reproduced.

As a result, the dense accumulation of states is a microscopic effect,
reflecting the importance of the content of the microscopic model space. 
Collective
models, geometric or algebraic, are not able to describe these states, implying
that the interpretation of $\beta$-bands or $\gamma$-bands are loosing their
geometric meaning with increasing energy, or requiring many local excited minima.

We also showed that the IBA has a serious defect with respect to the Pauli
principle, it does not take into account the substructure of the bosons as
fermion pairs. In the middle of the shell this turns out to be very important 
for the explanation of $B(E2)$-transition structure, as already shown
by K. T. Kecht \cite{Hecht-IBA}.

This contribution also demonstrates that in order to understand the basic characteristics of a spectrum and transitions, sometimes it is important 
to search first for a simple model Hamiltonian which carries the basic structure,
underpinning the more complicated and more realistic structure.
The most important part is the Hilbert space, i.e., the microscopic shell model
space. 
For a more complete description, however,
a realistic Hamiltonian and transition 
operators are required.

It would be interesting to repeat this inestigation using the
proxy-$SU(3)$ model \cite{proxy}, which takes into account also part of the
intruder levels.



\bmhead{Acknowledgements}

We acknowledge financial support form DGAPA-PAPIIT 
(IN100421).

\section*{Declarations}
All data generated or analyzed
during this study are included in
this published article.

\begin{appendices}

\FloatBarrier
\section{Parameter values and effective charges}
\label{app-Parameters}

\begin{center}
\begin{table}[h!]
\centering
\begin{tabular}{|c|c|c|c|c|c|c|}
\hline\hline
$^A_ZX_N$ &  $\chi$ & $a$ & $a_L$ & $b$ 
& $c$ & $d$ \\
\hline
$^{146}_{62}{\rm Sm}$ & 0. & 0.05016 & 0. & 6.15042 & 1.452~(-4) & $-4.99~(-5)$ 
\\
$^{148}_{62}{\rm Sm}$ & -0.03179 & 0.03404 & 0. & 0.3679 & $5.80494~(-5)$ 
& $2.9313~(-6)-006$  \\
$^{150}_{62}{\rm Sm}$ & -0.03024  & 0.03396 & 0.0015863 & 0.53909 &  1.2500~(-4) 
& $2.45~(-6)$ \\
$^{152)}_{62}{\rm Sm}$ & -0.14248 & 0.019541 & 0. & 1.036443 
& $7.67~(-5)$ & $6.77075~(-5)$  \\
$^{154}_{62}{\rm Sm}$ & -0.026920 & 0.012934 & 0. & 0.34326 & -2.4673~(-2) 
& -2.1378~(-4) \\
\hline
$^{154}_{64}{\rm Gd}$ & -0.02920 & 0.027614 & -0.0093864 & 0.17107 & $6.05~(-5)$  
& $1.32~(-6)$ \\
$^{156}_{64}{\rm Gd}$ & -0.081895 & 0.020407 & -0.0061690 & 0.25569 
& $5.00767~(-5)$  & $2.81191~(-5)$ \\
$^{158}_{64}{\rm Gd}$ & -0.040747 & 0.013441 & -0.00043394 & 0.27533 
& $4.81~(-5)$ & $8.81~(-6)$ \\
$^{160}_{64}{\rm Gd}$ & -0.029503 &  0.012242 & 0.0013702 & 0.23043 
& $2.41~(-5)$  & $7.42~(-7)$ \\
\hline 
$^{160}_{66}{\rm Dy}$ & -0.030616 & 0.014231 & -0.00013072 & 0.21886 
& $1.57~(-5)$ & $5.20~(-7)$ \\
$^{162}_{66}{\rm Dy}$ & -0.15613 & 0.028575 & -0.015353 & 0.19869 
& $1.86~(-5)$ & $4.77~{-5}$ \\
$^{164}_{66}{\rm Dy}$ & -0.029083 & 0.011095  & 0.00095592 & 0.17047 
& $2.17~(-5)$ & $-2.20~(-6)$ \\
\hline 
$^{162}_{68}{\rm Er}$ &-0.0273301907 & 0.01316 & 0.0025189 & 0.20157 
& $-8.36~(-6)$ & $-2.53~(-7)$ \\
$^{164}_{68}{\rm Er}$ & -0.03175476 & 0.0126808 & 0.0021365 & 0.19137 
& $1.15~(-5)$ & $1.89~(-6)$ \\
$^{166}_{68}{\rm Er}$ & -0.034697 & 0.013182 & 0. & 0.17498 & $2.30~(-5)$ 
& $1.05~(-6)$ \\
$^{168}_{68}{\rm Er}$ & -0.061823 &  0.012829 & 0.18475 & 0.00028826
& $6.94~(-5)$ & $-3.36~(-6)$ \\
$^{170}_{68}{\rm Er}$ & -0.012425 & 0.014336 & -0.0014601 & 0.344715155 
& $-2.16~(-5)$ & $1.37~(-5)$ \\
\hline
$^{168}_{70}{\rm Yb}$ & -0.047434 & 0.012372 & 0.0019079 & 0.22468 
& $3.34~(-5)$ & $-3.79~(-6)$ \\
$^{170}_{70}{\rm Yb}$ & -0.060589 & 0.0069394 & 0.0068854 & 0.30594 
& $3.37~(-5)$ & $1.27~(-5)$ \\
$^{172}_{70}{\rm Yb}$ & -0.029980' &  0.011692 & 0.0012167 & 0.27088 
& $1.66~(-5)$ & $3.98~(-6)$ \\
$^{174}_{70}{\rm Yb}$ & -0.029410 & 0.011426 & 0.00094857 & 0.29453 
& $9.57~(-6)$ & $1.77~(-6)$ \\
$^{176}_{70}{\rm Yb}$ & -0.023218 & 0.0012537 & 0.012061 & 0.38292 
& $6.75~(-7)$ & $8.56~(-7)$ \\
\hline
$^{174}_{72}{\rm Hf}$ & -0.035152 & 0.012429 & 0.0020678 & 0.32257 
& $-1.85~(-5)$ & $5.47~(-6)$ \\
$^{176}_{72}{\rm Hf}$ & -0.032171 & 0.012143 & 0.0017152 & 0.25675 
& $-5.86~(-6)$ & $2.24~(-6)$ \\
$^{178}_{72}{\rm Hf}$ & -0.037369 & 0.012148 & 0.0027187 & 0.33325 
& $-3.37~(-5)$ & $6.90~(-6)$ \\
$^{180}_{72}{\rm Hf}$ & -0.027891 & 0.012986 & 0.0023913 & 0.28192 
& $-2.51~(-5)$ & $-1.19~(-6)$ \\
$^{182}_{72}{\rm Hf}$ & -0.023513 & 0.013159 &  0.0029309 & 0.18036 
& $-1.98~(-5)$ & $1.46~(-5)$ \\
\hline
 \end{tabular}
\caption{Parameter values of the Hamiltonian, obtained via a fit.
}
\vspace{0.2cm}
\label{para}
\end{table}
\end{center}

\begin{center}
\begin{table}[h!]
\centering
\begin{tabular}{|c|c|c|c|c|}
\hline\hline
$^{146}_{62}$Sm & $^{148}_{62}$Sm & $^{150}_{62}$Sm 
& $^{152}_{62}$Sm & $^{154}_{62}$Sm  \\
\hline
0.42341 & 0.31750 & 0.24003 & 0.48759 & 0.45896 \\
\hline
\hline
$^{154}_{64}$Gd & $^{156}_{64}$Gd & $^{158}_{64}$Gd 
& $^{160}_{64}$Gd &  \\
\hline
0.46128 & 0.47790 & 0.49159 & 0.46825 &  \\
\hline
\hline
$^{160}_{66}$Dy & $^{162}_{66}$Dy & $^{164}_{66}$Dy & &  \\
\hline
0.49126 & 0.47522 & 0.48536 & & \\
\hline
\hline 
$^{162}_{68}$Er & $^{164}_{68}$Er & $^{166}_{68}$Er & 
$^{168}_{68}$Er & $^{170}_{68}$Er \\
\hline
0.48469 & 0.47069 & 0.49455 & 0.50607 & 0.47474 \\
\hline
\hline
$^{168}_{70}$Yb & $^{170}_{70}$Yb 
& $^{172}_{70}$Yb 
& $^{174}_{70}$Yb & $^{176}_{70}$Yb \\
\hline
0.48776 & 0.49389 & 0.49479 & 0.48392 & 0.45122 \\
\hline
\hline
$^{174}_{72}$Hf & $^{176}_{72}$Hf & $^{178}_{72}$Hf 
& $^{180}_{72}$Hf & $^{182}_{72}$Hf \\
\hline
0.44589 & 0.47313 & 0.47510 & 0.41743 & 0.51564 \\
\hline
\hline
 \end{tabular}
\caption{
List of the effective charges assigned to the nuclei analyzed in this paper. All obtained values are approximately 0.5.
}
\vspace{0.2cm}
\label{e-eff}
\end{table}
\end{center}

\begin{center}
\begin{table}[h!]
\centering
\begin{tabular}{|c|c|c|c|c|c|c|}
\hline\hline
$^A_ZX_N$ &  $\epsilon_2$ & $(\tilde{Z},\tilde{N})$ & $(Z_n,N_n)$ & 
$(\tilde{\lambda}_\pi,\tilde{\mu}_\pi)$ 
& $(\tilde{\lambda}_\nu,\tilde{\mu}_\nu)$ & 
$(\tilde{\lambda} , \tilde{\mu} )$ \\
\hline
$^{146}_{62}{\rm Sm}_{84}$ & 0. & [26,40] & [8,2] & (10,4) & (8,0) & (18,4) \\
$^{148}_{62}{\rm Sm}_{86}$ & 0.16 & [26,42] & [8,4] & (10,4) & (12,2) & (22,6) \\
$^{150}_{62}{\rm Sm}_{88}$ & 0.19 & [26,44] & [8,6] & (10,4) & (18,0) & (28,4) \\
$^{152}_{62}{\rm Sm}_{90}$ & 0.22 & [26,44] & [8,6] & (10,4) & (18,0) & (28,4) \\
$^{154}_{62}{\rm Sm}_{92}$ & 0.25 & [26,46] & [8,8] & (10,4) & (18,4) & (28,8) \\
\hline
$^{154}_{64}{\rm Gd}_{90}$ & 0.22 & [26,44] & [8,6] & (10,4) & (18,0) & (28,8) \\
$^{156}_{64}{\rm Gd}_{92}$ & 0.24 & [26,46] & [8,8] & (10,4) & (18,4) & (28,8) \\
$^{158}_{64}{\rm Gd}_{94}$ & 0.26 & [26,46] & [8,8] & (10,4) & (18,4) & (28,8) \\
$^{160}_{64}{\rm Gd}_{96}$ & 0.26 & [26,48] & [8,10] & (10,4) & (20,4) & (30,8) \\
\hline
$^{160}_{66}{\rm Dy}_{94}$ & 0.25 & [28,46] & [10,8] & (10,4) & (18,4) & (28,8) \\
$^{162}_{66}{\rm Dy}_{96}$ & 0.26 & [28,48] & [10,10] & (10,4) & (20,4) & (30,8) \\
$^{164}_{66}{\rm Dy}_{98}$ & 0.27 & [28,48] & [10,10] & (10,4) & (20,4) & (30,8) \\
\hline 
$^{162}_{68}{\rm Er}_{94}$ & 0.25 & [28,46] & [10,8] & (10,4) & (18,4) & (28,8) \\
$^{164}_{68}{\rm Er}_{96}$ & 0.26 & [28,48] & [10,10] & (10,4) & (20,4) & (30,8) \\
$^{166}_{68}{\rm Er}_{98}$ & 0.26 & [28,48] & [10,10] & (10,4) & (20,4) & (30,8) \\
$^{168}_{68}{\rm Er}_{100}$ & 0.27 & [28,50] & [10,12] & (10,4) & (24,0) & (34,4) \\
$^{170}_{68}{\rm Er}_{102}$ & 0.27 & [28,52] & [10,14] & (10,4) & (20,6) & (30,10) \\
\hline
$^{168}_{70}{\rm Yb}_{98}$ & 0.28 & [30,48] & [12,10] & (4,10) & (20,4) & (32,4) \\
$^{170}_{70}{\rm Yb}_{100}$ & 0.26 & [30,50] & [12,12] & (4,10) & (24,0) & (36,0) \\
$^{172}_{70}{\rm Yb}_{102}$ & 0.27 & [30,52] & [12,14] & (4,10) & (20,6) & (12,28) \\
$^{174}_{70}{\rm Yb}_{104}$ & 0.26 & [30,52] & [12,14] & (4,10) & (20,6) & (12,28) \\
$^{176}_{70}{\rm Yb}_{106}$ & 0.26 & [30,54] & [12,16] & (4,10) & (6,20) & (10,30) \\
\hline
$^{174}_{72}{\rm Hf}_{102}$ & 0.26 & [32,52] & [14,14] & (0,12) & (20,6) & (8,30) \\
$^{176}_{72}{\rm Hf}_{104}$ & 0.25 & [32,52] & [14,14] & (0,12) & (20,6) & (8,30) \\
$^{178}_{72}{\rm Hf}_{106}$ & 0.25 & [32,54] & [14,16] & (0,12) & (6,20) & (6,32) \\
$^{180}_{72}{\rm Hf}_{108}$ & 0.24 & [32,54] & [14,16] & (0,12) & (6,20) & (6,32) \\
$^{182}_{72}{\rm Hf}_{110}$ & 0.24 & [32,56] & [14,18] & (0,12) & (0,24) & (0,36) \\
\hline
 \end{tabular}
\caption{
$\widetilde{SU}(3)$ data for the nuclei considered in this contribution. $\epsilon_2$ is the Nilsson quadrupole deformation, $Z_n$ and $N_n$ refer to the 
normal number of protons and neutrons.
$\tilde{Z}$ and $\tilde{N}$ are the number of normal protons, in the $\tilde{\eta}=3$ pseudo-shell, and neutrons, in the $\tilde{\eta}=4$ pseudo-shell. 
$(\tilde{\lambda}_\pi,\tilde{\mu}_\pi)$ and
$(\tilde{\lambda}_\nu,\tilde{\mu}_\nu)$ are the lowest 
$\widetilde{SU}(3)$ irreps in the proton and neutron
part, respectively. Finally, 
$(\tilde{\lambda} , \tilde{\mu} )$ is the lowest lying total $\widetilde{SU}(3)$
irrep.
}
\vspace{0.2cm}
\label{tab-nor}
\end{table}
\end{center}

\FloatBarrier
\section{Quantum numbers and $\widetilde{SU}(3)$ model space}
\label{appQN}

\begin{center}
\begin{table}[h!]
\centering
\begin{tabular}{|c|c|c|c|c|}
\hline\hline
$^{146}_{62}$Sm:  & $^{148}_{62}$Sm:   &
$^{150}_{62}$Sm:  & $^{152}_{62}$Sm: 
& $^{154}_{62}$Sm: \\ 
\hline
\hline
(15,4)$_3$ & (19,6)$_5$  & (20,11)$_2$ & (20,11)$_2$ & (24,10)$_8$ \\
(11,9)$_1$ & (15,11)$_1$ & (24,6)$_9$ & (24,6)$_9$ & (30,1)$_2$ \\
(8,12)$_1$ & (18,8)$_5$  & (23,8)$_3$ & (23,8)$_3$ & (29,3)$_4$ \\
(14,6)$_4$ & (23,1)$_1$  & (28,1)$_1$ & (28,1)$_1$ & (28,5)$_4$ \\
13,8)$_1$ & (22,3)$_1$  & (22,10)$_1$ & (22,10)$_1$ & (23,12)$_2$ \\
(18,1)$_1$ & (17,10)$_1$ & (27,3)$_3$ & (27,3)$_3$ & (27,7)$_4$ \\
(17,3)$_2$ & (21,5)$_3$  & (26,5)$_1$ & (26,5)$_1$ & (26,9)$_2$ \\
(16,5)$_1$ & (20,7)$_1$  & (25,7)$_1$ & (25,7)$_1$ & (32,0)$_1$ \\
(20,0)$_1$ & (24,2)$_1$  & (30,0)$_1$ & (30,0)$_1$ & (30,4)$_2$ \\
(18,4)$_1$ & (22,6)$_1$  & (28,4)$_1$ & (28,4)$_1$ & (28,8)$_1$ \\
\hline
 \end{tabular}
\caption{The 10 largest
$\widetilde{SU}(3)$ irreps for the Sm-isotopes.
The multiple appearance (multiplicity) of each irrep is indicated by the sub-index
next to each irrep.
} 
\vspace{0.2cm}
\label{Sm-irrep}
\end{table}
\end{center}

\begin{center}
\begin{table}[h!]
\centering
\begin{tabular}{|c|c|c|c|}
\hline\hline
$^{154}_{64}$Gd:  & $^{156}_{64}$Gd:  &
$^{158}_{64}$Gd:  & $^{160}_{64}$Gd: 
\\
\hline
\hline
(20,11)$_2$ & (24,10)$_8$  & (24,10)$_8$ & (31,3)$_4$ \\
(24,6)$_9$ & (30,1)$_2$ & (30,1)$_2$ & (25,12)$_4$\\
(23,8)$_3$ & (29,3)$_4$  & (29,3)$_4$ & (30,5)$_5$  \\
(28,1)$_1$ & (28,5)$_4$  & (28,5)$_4$ & (29,7)$_5$  \\
(22,10)$_1$ & (23,12)$_2$  & (23,12)$_2$ & (24,14)$_1$ \\
(27,3)$_3$ & (27,7)$_4$ & (27,7)$_4$ & (28,9)$_2$  \\
(26,5)$_1$ & (26,9)$_2$  & (26,9)$_2$ & (34,0)$_1$  \\
(25,7)$_1$ & (32,0)$_1$  & (32,0)$_1$ & (27,11)$_1$  \\
(30,0)$_1$ & (30,4)$_2$  & (30,4)$_2$ & (32,4)$_2$ \\
(28,4)$_1$ & (28,8)$_1$  & (28,8)$_1$ & (30,8)$_1$ \\
\hline
 \end{tabular}
\caption{The 10 largest
$\widetilde{SU}(3)$ irreps for the Gd-isotopes.
The multiple appearance (multiplicity) of each irrep is indicated by the sub-index
next to each irrep.
} 
\vspace{0.2cm}
\label{Gd-irrep}
\end{table}
\end{center}

\begin{center}
\begin{table}[h!]
\centering
\begin{tabular}{|c|c|c|}
\hline\hline
$^{160}_{66}$Dy:   & $^{162}_{66}$Dy:   &
$^{164}_{66}$Dy:   
\\
\hline
\hline
(29,3)$_4$ & (25,12)$_5$  & (25,12)$_5$ \\
(28,5)$_5$ & (18,20)$_1$ & (18,20)$_1$   \\
(23,12)$_3$ & (21,17)$_2$  & (21,17)$_2$   \\
(27,7)$_5$ & (29,7)$_6$  & (29,7)$_6$   \\
(22,14)$_1$ & (24,14)$_3$  & (24,14)$_3$  \\
(26,9)$_2$ & (28,9)$_2$ & (28,9)$_2$   \\
(32,0)$_1$ & (34,0)$_1$  & (34,0)$_1$  \\
(25,11)$_1$ & (27,11)$_2$  & (27,11)$_2$   \\
(30,4)$_2$ & (32,4)$_2$  & (32,4)$_2$   \\
(28,8)$_1$ & (30,8)$_1$  & (30,8)$_1$ \\
\hline
 \end{tabular}
\caption{The 10 largest
$\widetilde{SU}(3)$ irreps for the Dy-isotopes.
The multiple appearance (multiplicity) of each irrep is indicated by the sub-index
next to each irrep.
}
\vspace{0.2cm}
\label{Dy-irrep}
\end{table}
\end{center}

\begin{center}
\begin{table}[h!]
\centering
\begin{tabular}{|c|c|c|c|c|}
\hline\hline
$^{162}_{68}$Er:  & $^{164}_{68}$Er:  &
$^{166}_{68}$Er: & $^{168}_{68}$Er:
& $^{170}_{68}$Er: \\
\hline
\hline
(29,3)$_4$ & (25,12)$_5$  & (25,12)$_5$ & (33,3)$_3$ & (21,19)$_4$ \\
(28,5)$_5$ & (18,20)$_1$ & (18,20)$_1$ & (27,12)$_2$ & (18,22)$_4$ \\
(23,12)$_3$ & (21,17)$_2$  & (21,17)$_2$ & (32,5)$_2$ & (28,11)$_3$ \\
(27,7)$_5$ & (29,7)$_6$  & (29,7)$_6$ & (20,20)$_1$ & (24,16)$_4$ \\
(22,14)$_1$ & (24,14)$_3$  & (24,14)$_3$ & (23,17)$_1$ & (34,2)$_1$ \\
(26,9)$_2$ & (28,9)$_2$ & (28,9)$_2$ & (31,7)$_2$ & (15,25)$_2$ \\
(32,0)$_1$ & (34,0)$_1$  & (34,0)$_1$ & (26,14)$_1$ & (32,6)$_2$ \\
(25,11)$_1$ & (27,11)$_2$  & (27,11)$_2$ & (30,9)$_1$ & (27,13)$_2$ \\
(30,4)$_2$ & (32,4)$_2$  & (32,4)$_2$ & (36,0)$_1$ & (12,28)$_1$ \\
(28,8)$_1$ & (30,8)$_1$  & (30,8)$_1$ & (34,4)$_1$ & (30,10)$_1$ \\
\hline
 \end{tabular}
\caption{The 10 largest
$\widetilde{SU}(3)$ irreps for the Er-isotopes.
The multiple appearance (multiplicity) of each irrep is indicated by the sub-index
next to each irrep.
}
\vspace{0.2cm}
\label{Er-irrep}
\end{table}
\end{center}

\begin{center}
\begin{table}[h!]
\centering
\begin{tabular}{|c|c|c|c|c|}
\hline\hline
$^{168}_{70}$Yb:   & $^{170}_{70}$Yb:  &
$^{172}_{70}$Yb: & $^{174}_{70}$Yb: 
& $^{176}_{70}$Yb: \\
\hline
\hline
(31,3)$_3$ & (18,21)$_1$  & (13,26)$_3$ & (13,26)$_3$ & (26,13)$_1$ \\
(25,12)$_2$ & (21,18)$_2$ & (8,30)$_3$ & (8,30)$_3$ & (30,8)$_1$ \\
(30,5)$_2$ & (10,28)$_1$  & (21,19)$_1$ & (21,19)$_1$ & (19,21)$_1$ \\
(18,20)$_1$ & (28,10)$_3$  & (18,22)$_2$ & (18,22)$_2$ & (22,18)$_1$ \\
(29,7)$_2$ & (15,24)$_1$  & (28,11)$_1$ & (28,11)$_1$ & (11,28)$_3$ \\
(21,17)$_1$ & (24,15)$_2$ & (24,16)$_1$ & (24,16)$_1$ & (16,24)$_2$ \\
(24,14)$_1$ & (33,3)$_2$  & (34,2)$_1$ & (34,2)$_1$ & (2,34)$_1$ \\
(28,9)$_1$ & (32,5)$_2$  & (15,25)$_1$ & (15,25)$_1$ & (6,32)$_2$ \\
(34,4)$_1$ & (20,20)$_1$  & (32,6)$_1$ & (32,6)$_1$ & (13,27)$_1$ \\
(32,4)$_1$ & (36,0)$_1$  & (12,28)$_1$ & (12,28)$_1$ & (10,30)$_1$ \\
\hline
 \end{tabular}
\caption{The 10 largest
$\widetilde{SU}(3)$ irreps for the Yb-isotopes.
The multiple appearance (multiplicity) of each irrep is indicated by the sub-index
next to each irrep.
}
\vspace{0.2cm}
\label{Yb-irrep}
\end{table}
\end{center}

\begin{center}
\begin{table}[h!]
\centering
\begin{tabular}{|c|c|c|c|c|}
\hline\hline
$^{174}_{72}$Hf: & $^{176}_{72}$Hf:  &
$^{178}_{72}$Hf: & $^{180}_{72}$Hf:
& $^{170}_{72}$Hf: \\
\hline
\hline
(7,29)$_6$ & (7,29)$_6$  & (15,23)$_4$ & (15,23)$_4$ & (2,32)$_8$ \\
(17,21)$_4$ & (17,21)$_4$ & (24,14)$_1$ & (24,14)$_1$ & (4,31)$_4$ \\
(23,15)$_2$ & (23,15)$_2$  & (1,33)$_2$ & (1,33)$_2$ & (16,22)$_1$ \\
(14,24)$_4$ & (14,24)$_4$  & (3,32)$_3$ & (2,32)$_3$ & (6,30)$_6$ \\
(9,28)$_2$ & (9,28)$_2$  & (12,26)$_4$ & (12,26)$_4$ & (13,25)$_1$ \\
(26,12)$_1$ & (26,12)$_1$ & (5,31)$_4$ & (5,31)$_4$ & (8,29)$_2$ \\
(0,34)$_1$ & (0,34)$_1$  & (7,30)$_2$ & (7,30)$_2$ & (10,28)$_1$ \\
(11,27)$_2$ & (11,27)$_2$  & (9,29)$_2$ & (9,29)$_2$ & (3,33)$_2$ \\
(4,32)$_2$ & (4,32)$_2$  & (2,34)$_1$ & (2,34)$_1$ & (5,32)$_1$ \\
(8,30)$_1$ & (8,30)$_1$  & (6,32)$_1$ & (6,32)$_1$ & (0,36)$_1$ \\
\hline
 \end{tabular}
\caption{The 10 largest
$\widetilde{SU}(3)$ irreps for the Hf-isotopes.
The multiple appearance (multiplicity) of each irrep is indicated by the sub-index
next to each irrep.
} 
\vspace{0.2cm}
\label{Hf-irrep}
\end{table}
\end{center}

\FloatBarrier
\section{Energies}
\label{appenergies}

\begin{center}
\begin{table}[h!]
\centering
\begin{tabular}{|c|c|c|c|c|c|}
\hline\hline
$J^P_i$ & $^{146}_{62}$Sm  & $^{148}_{62}$Sm & $^{150}_{62}$Sm & $^{152}_{62}$Sm
& $^{154}_{62}$Sm\\
\hline
\hline
$0^+_1$ & 0. (0.) & 0. (0.) & 0. (0.) & 0. (0.) & 0. (0.) \\
$2^+_1$ & 0.301 (0.747) & 0.193 (0.550) & 0.213 (0.334) & 0.117 (0.122) & 0.078 (0.082) \\
$4^+_1$ & 1.003 (1.381) & 0.641 (1.180) & 0.711 (0.773) & 0.391 (0.366) & 0.259 (0.267) \\ 
$6^+_1$ & 2.107 (1.812) & 1.346 (1.279) & 1.493 (1.279) & 0.821 (0.810) & 0.543 (0.544) \\
$2^+_2$ & 1.482 (1.648) & 1.296 (1.454) & 0.977 (1.046) & 0.843 (1.293) & 1.192 (1.178) \\
$3^+_1$ & 2.348 (2.270) & 1.904 (1.904) & 1.529 (1.505) & 1.307 (1.233) & 1.528 (1.539) \\
$4^+_2$ & 2.184 (2.281) & 1.744 (1.733) & 1.475 (1.449) & 1.116 (1.023) & 1.373 (1.338) \\
$5^+_1$ & 3.251 (2.898) & 2.553 (2.148) & 2.112 (1.883) & 1.659 (1.560) & 1.762 (1.805) \\
$0^+_2$ & 1.181 (2.211) & 1.104 (1.424) & 0.764 (0.740) & 0.726 (0.685) & 1.114 (1.151*) \\
$2^+_3$ & 1.931 (2.156) & 1.664 (1.664) & 1.353 (1.194) & 1.233 (1.234) & 1.438 (1.286) \\
\hline
\end{tabular}
\caption{
Energies of the Sm isotopes. The theoretical values are listed for each nucleus 
in the first row of each column, 
while the experimental values are in the parenthesis. 
A star (*) indicates that the theory predicts a degeneracy of irreps. In this case, for the experimental values an average, estimated values are given.
The experimental values, except for the $0^+$-states, are taken from \cite{brook}.
The information on the $0^+$-states are taken from \cite{ani2025}.
}
\vspace{0.2cm}
\label{Sm-energy}
\end{table}
\end{center}

\begin{center}
\begin{table}[h!]
\centering
\begin{tabular}{|c|c|c|c|c|c|}
\hline\hline
$J^P_i$ & $^{146}_{62}$Sm  & $^{148}_{62}$Sm & $^{150}_{62}$Sm & $^{152}_{62}$Sm
& $^{154}_{62}$Sm\\
\hline
\hline
$0^+_1$ & 0. (0.) & 0. (0.) & 0. (0.) & 0. (0.) & 0. (0.) \\
$0^+_2$ & 1.181 (2.211) & 1.104 (1.424)$_B$ & 0.764 (0.740) & 0.726 (0.685) & 1.114 (1.151*) \\
$0^+_3$ & 1.630 (2.650*) & 2.161 (2.161*)$_B$ & 1.645 (1.256) & 1.116 (1.083) & 1.136 (-) \\
$0^+_4$ & 2.063 (-) & 4.788 (-) & 2.556 (2.266*) & 2.018 (2.000*) & - (-) \\
\hline
 \end{tabular}
\caption{
$0^+$ spectrum of the Sm isotopes.
A star (*) indicates that the theory predicts a degeneracy of irreps. In this case, for the
experimental values an average, estimated values are given. The $0^+$ data are taken exclusively from \cite{ani2025}.
}
\vspace{0.2cm}
\label{Sm-0}
\end{table}
\end{center}

\begin{center}
\begin{table}[h!]
\centering
\begin{tabular}{|c|c|c|c|c|}
\hline\hline
$J^P_i$ & $^{154}_{64}$Gd  & $^{156}_{64}$Gd & $^{158}_{64}$Gd & $^{160}_{64}$Gd \\
\hline
\hline
$0^+_1$ & 0. (0.) de nuevo & 0. (0.) de nuevo & 0. (0.) & 0. (0.)\\
$2^+_1$ & 0.109 (0.123) & 0.085 (0.089) & 0.078 (0.080) & 0.082 (0.075) \\
$4^+_1$ & 0.365 (0.371) & 0.285 (0.288) & 0.260 (0.261) & 0.272 (0.249) \\
$6^+_1$ & 0.766 (0.718) & 0.598 (0.585) & 0.546 (0.539) & 0.572 (0.515) \\
$2^+_2$ & 0.793 (0.815) & 1.108 (1.129) & 1.179 (1.187) & 1.003 (0.989) \\
$3^+_1$ & 1.128 (1.128) & 1.249 (1.248) & 1.268 (1.266) & 1.052 (1.057) \\
$4^+_2$ & 1.048 (1.048) & 1.308 (1.298) & 1.361 (1.358) & 1.194 (1.070) \\
$5^+_1$ & 1.794 (1.433) & 1.727 (1.507) & 1.518 (1.481) & 1.248 (1.173) \\
$0^+_2$ & 0.684 (0.681) & 1.125 (1.109*) & 1.260 (1.251*) & 1.374 (1.380*) \\
$2^+_3$ & 0.794 (0.996) & 1.210 (1.154) & 1.238 (1.260) & 1.403 (1.436) \\
\hline
 \end{tabular}
\caption{
Energies of the Gd isotopes. The theoretical values are listed for each nucleus 
in the first row of each column, while the experimental values are in the parenthesis. 
A star (*) indicates that the theory predicts a degeneracy of irreps. In this case, for the
experimental values an average, estimated values are given.
The experimental values, except for the $0^+$-states, are taken from \cite{brook}.
The information on the $0^+$-states are taken from \cite{ani2025}
}
\vspace{0.2cm}
\label{Gd-energy}
\end{table}
\end{center}

\begin{center}
\begin{table}[h!]
\centering
\begin{tabular}{|c|c|c|c|c|}
\hline\hline
$J^P_i$ & $^{154}_{64}$Gd  & $^{156}_{64}$Gd & $^{158}_{64}$Gd & $^{160}_{64}$Gd
\\
\hline
\hline
$0^+_1$ & 0. (0.) & 0. (0.) & 0. (0.) & 0. (0.) \\
$0^+_2$ & 0.684 (0.681) & 1.125 (1.109*) & 1.260 (1.251) & 1.375  (1.469*)\\
$0^+_3$ & 1.178 (1.182) & 1.637 (1.715) & 1.809 (1.743) & 1.667 (1.558) \\
$0^+_4$ & 3.188 (1.808*) & 2.111 (1.974*) & 2.363 (2.365) & 1.964 (-) \\
\hline
 \end{tabular}
\caption{
$0^+$ spectrum of the Gd isotopes.
A star (*) indicates that the theory predicts a degeneracy of irreps. In this case, for the
experimental values an average, estimated values are given. The $0^+$ data are taken exclusively from \cite{ani2025}.
}
\vspace{0.2cm}
\label{Gd-0}
\end{table}
\end{center}

\begin{center}
\begin{table}[h!]
\centering
\begin{tabular}{|c|c|c|c|}
\hline\hline
$J^P_i$ & $^{160}_{66}$Dy  & $^{162}_{66}$Dy & $^{164}_{66}$Dy \\
\hline
\hline
$0^+_1$ & 0. (0.) de nuevo & 0. (0.) & 0. (0.) \\
$2^+_1$ & 0.085 (0.087) & 0.079 (0.081) & 0.072 (0.073) \\
$4^+_1$ & 0.282 (0.284) & 0.264 (0.266) & 0.241 (0.242) \\
$6^+_1$ & 0.592 (0.581) & 0.555 (0.549) & 0.506 (0.501) \\
$2^+_2$ & 0.960 (0.966) & 0.874 (0.888) & 0.754 (0.762) \\
$3^+_1$ & 1.048 (1.049) & 1.322 (0.963) & 0.804 (0.828) \\
$4^+_2$ & 1.157 (1.156) & 1.059 (1.161) & 0.923 (0.916) \\
$5^+_1$ & 1.306 (1.289) & 1.107 (1.183) & 0.986 (1.025) \\
$0^+_2$ & 0.818 (0.370*) & 1.555 (1.533) & 1.655 (1.655) \\
$2^+_3$ & 1.453 (1.350) & 1.608 (1.453) & 1.728 (1.716) \\
\hline
 \end{tabular}
\caption{
Energies of the Dy isotopes. The theoretical values are listed for each nucleus 
in the first row of each column, while the experimental values are in the parenthesis. 
A star (*) indicates that the theory predicts a degeneracy of irreps. In this case, for the
experimental values an average, estimated values are given.
The experimental values, except for the $0^+$-states, are taken from \cite{brook}.
The information on the $0^+$-states are taken from \cite{ani2025}.
}
\vspace{0.2cm}
\label{Dy-energy}
\end{table}
\end{center}

\begin{center}
\begin{table}[h!]
\centering
\begin{tabular}{|c|c|c|c|}
\hline\hline
$J^P_i$ & $^{160}_{66}$Dy  & $^{162}_{66}$Dy & $^{164}_{66}$Dy \\
\hline
\hline
$0^+_1$ & 0. (0.) & 0. (0.) & 0. (0.)  \\
$0^+_2$ & 1.368 (1.378*) & 1.528 (1.533*) & 1.655 (1.720*) \\
$0^+_3$ & 1.889 (1.708) & 2.235 (1.815) & 2.332 (-) \\
$0^+_4$ & 1.955 (1952) & 2.603 (2.316*) & 2.362 (-) \\
$0^+_5$ & 4.039 (-)& 2.783 (2.588)$_B$ & 2.440 (-) \\
\hline
 \end{tabular}
\caption{
$0^+$ spectrum of the Dy isotopes.
A star (*) indicates that the theory predicts a degeneracy of irreps. In this case, for the
experimental values an average, estimated values are given. The $0^+$ data are taken exclusively from \cite{ani2025}.
} 
\vspace{0.2cm}
\label{Dy-0}
\end{table}
\end{center}

\begin{center}
\begin{table}[h!]
\centering
\begin{tabular}{|c|c|c|c|c|c|}
\hline\hline
$J^P_i$ & $^{162}_{68}$Er  & $^{164}_{68}$Er & $^{166}_{68}$Er & $^{168}_{68}$Er
& $^{170}_{68}$Er\\
\hline
\hline
$0^+_1$ & 0. (0.) & 0. (0.) & 0. (0.) & 0. (0.) & 0. (0.) \\
$2^+_1$ & 0.094 (0.102) & 0.089 (0.091) & 0.079 (0.081) & 0.079 (0.080) & 0.077 (0.079) \\
$4^+_1$ & 0.314 (0.330) & 0.296 (0.299) & 0.264 (0.265) & 0.262 (0.264) & 0.258 (0.260) \\
$6^+_1$ & 0.658 (0.667) & 0.622 (0.614) & 0.554 (0.545) & 0.551 (0.549) & 0.541 (0.541) \\
$2^+_2$ & 0.900 (0.901) & 0.854 (0.860) & 0.779 (0.786) & 0.818 (0.821) & 0.954 (0.934) \\
$3^+_1$ & 0.934 (1.002) & 0.892 (0.946) & 0.858 (0.859) & 0.890 (0.896) & 1.096 (1.011) \\
$4^+_2$ & 1.120 (1.128) & 1.062 (1.058) & 0.964 (0.956) & 1.001 (0.995) & 1.134 (1.113) \\
$5^+_1$ & 1.125 (1.286) & 1.082 (1.197) & 1.095 (1.075) & 1.115 (1.118) & 1.380 (1.237) \\
$0^+_2$ & 1.102 (1.087) & 1.231  (1.332*) & 1.549 (1.590*) & 1.217 (1.217) & 0.876 (0.891) \\
$2^+_3$ & 1.196 (1.171) & 1.320 (1.314) & 1.618 (1.528) & 1.296 (1.276) & 0.984 (0.960) \\
\hline
 \end{tabular}
\caption{
Energies of the Er isotopes. The theoretical values are listed for each nucleus 
in the first row of each column, while the experimental values are in the parenthesis. 
A star (*) indicates that the theory predicts a degeneracy of irreps. In this case, for the average, estimated values are given.
The experimental values, except for the $0^+$-states, are taken from \cite{brook}.
The information on the $0^+$-states are taken from \cite{ani2025}.
}
\vspace{0.2cm}
\label{Er-energy}
\end{table}
\end{center}

\begin{center}
\begin{table}[h!]
\centering
\begin{tabular}{|c|c|c|c|c|c|}
\hline\hline
$J^P_i$ & $^{162}_{68}$Er  & $^{164}_{68}$Er & $^{166}_{68}$Er & $^{168}_{68}$Er
& $^{170}_{68}$Er \\
\hline
\hline
$0^+_1$ & 0. (0.) & 0. (0.) & 0. (0.) & 0. (0.) & 0. (0.) \\
$0^+_2$ & 1.102 (1.087*) & 1.231 (1.332*) & 1.549 (1.590*) & 1.217 (1.217) & 0.877 (0.891) \\
$0^+_3$ & 1.574 (2.114) & 1.761 (1.702) & 1.917 (1.934) & 1.434 (1.422) & 1.384 (1.324*) \\
$0^+_4$ & 2.654 (-) & 2.189 (1.970*) & 2.135 (2.100*) & 2.161 (1.834) & 2.122  (-) \\
$0^+_5$ & 4.704 (-) & 2.277 (-) & 2.214 () & 2.782 (2.780+) & 2.517 (-) \\
\hline
 \end{tabular}
\caption{
$0^+$ spectrum of the Er isotopes.
A star (*) indicates that the theory predicts a degeneracy of irreps. In this case, for the
experimental values an average, estimated values are given. The $0^+$ data are taken exclusively from \cite{ani2025}.
}
\vspace{0.2cm}
\label{Er-0}
\end{table}
\end{center}

\begin{center}
\begin{table}[h!]
\centering
\begin{tabular}{|c|c|c|c|c|c|}
\hline\hline
$J^P_i$ & $^{168}_{70}$Yb  & $^{170}_{70}$Yb & $^{172}_{70}$Yb & $^{174}_{70}$Yb
& $^{176}_{70}$Yb\\
\hline
\hline
$0^+_1$ & 0. (0.) & 0. (0.) & 0. (0.) & 0. (0.) & 0. (0.) \\
$2^+_1$ & 0.086 (0.088) & 0.083 (0.084) & 0.077 (0.079) & 0.074 (0.076) & 0.080 (0.082) \\
$4^+_1$ & 0.286 (0.287) & 0.276 (0.277) & 0.258 (0.260) & 0.247 (0.253) & 0.266 (0.272) \\
$6^+_1$ & 0.600 (0.585) & 0.581 (0.573) & 0.542 (0.540) & 0.520 (0.526) & 0.559 (0.565) \\
$2^+_2$ & 0.984 (0.984) & 1.058 (1.138) & 1.114 (1.118) & 1.252 (1.561) & 1.197 (1.200) \\
$3^+_1$ & 1.024 (1.067) & 1.224 (1.225) & 1.209 (1.172) & 1.304 (1.606) & 1.331 (1.336) \\
$4^+_2$ & 1.184 (1.171) & 1.251 (1.292) & 1.295 (1.263) & 1.426 (1.702) & 1.383 (1.341) \\
$5^+_1$ & 1.213 (1.302) & 1.225 (1.460) & 1.398 (1.376) & 1.492 (1.820) & 1.136 (1.492) \\
$0^+_2$ & 1.155 (1.155) & 0.975 (1.069) & 1.036 (1.043) & 1.491 (1.487) & 1.117 (1.778*) \\
$2^+_3$ & 1.240 (1.233) & 1.263 (1.145) & 1.161 (1.466) & 1.548 (1.634) & 1.540 (1.261) \\
\hline
 \end{tabular}
\caption{
Energies of the Yb isotopes. The theoretical values are listed for each nucleus 
in the first row of each column, while the experimental values are in the parenthesis. 
A star (*) indicates that the theory predicts a degeneracy of irreps. 
In this case, the values listed are average and estimated values rather than precise measurements. 
The experimental values, except for the $0^+$-states, are taken from \cite{brook}.
The information on the $0^+$-states are taken from \cite{ani2025}.
}
\vspace{0.2cm}
\label{Yb-energy}
\end{table}
\end{center}

\begin{center}
\begin{table}[h!]
\centering
\begin{tabular}{|c|c|c|c|c|c|}
\hline\hline
$J^P_i$ & $^{168}_{70}$Yb  & $^{170}_{70}$Yb & $^{172}_{70}$Yb & $^{174}_{70}$Yb
& $^{176}_{70}$Yb \\
\hline
\hline
$0^+_1$ & 0. (0.) & 0. (0.) & 0. (0.) & 0. (0.) & 0. (0.) \\
$0^+_2$ & 1.155 (1.155) & 0.975 (1.069) & 1.036 (1.043)$_B$ & 1.491 (1.487) 
& 1.117  (1.778*)$_B$ \\
$0^+_3$ & 1.340 (1.197) & 1.180(1.425+) & 1.714 (1.405)$_B$ & 1.894 (1.886) 
& 1.740 (-)$_B$ \\
$0^+_4$ & 1.414 (1.543) & 1.868 (2.088)& 1.925 (1.794)$_B$ & 1.924 (-)
& 1.763 (-) \\
$0^+_5$ & 2.575 (-)& 2.338 (2.210*) & 2.009 (1.895)$_B$ & 2.081 (-) & 2.066 (-) \\
$0^+_6$ & 6.321 (-) & 3.501 (2.709*) & 2.479 (-) & 2.695 (-) & 2.233 (-) \\
\hline
 \end{tabular}
\caption{
$0^+$ spectrum of the Yb isotopes.
A star (*) indicates that the theory predicts a degeneracy of irreps. In this case, for the
experimental values an average, estimated values are given. The $0^+$ data are taken exclusively from \cite{ani2025}.
}
\vspace{0.2cm}
\label{Yb-0}
\end{table}
\end{center}

\begin{center}
\begin{table}[h!]
\centering
\begin{tabular}{|c|c|c|c|c|c|}
\hline\hline
$J^P_i$ & $^{174}_{72}$Hf  & $^{176}_{72}$Hf & $^{178}_{72}$Hf & $^{180}_{72}$Hf
& $^{182}_{72}$Hf\\
\hline
\hline
$0^+_1$ & 0. (0.) & 0. (0.) & 0. (0.) & 0. (0.) & 0. (0.) \\
$2^+_1$ & 0.087 (0.091) & 0.083 (0.088) & 0.091 (0.093) & 0.092 (0.093) & 0.097 (0.098) \\
$4^+_1$ & 0.290 (0.297) & 0.277 (0.290) & 0.303 (0.307) & 0.308 (0.309) & 0.322 (0.322) \\
$6^+_1$ & 0.609 (0.608) & 0.582 (0.597) & 0.637 (0.632) & 0.646 (0.641) & 0.676 (0.666) \\
$2^+_2$ & 0.900 (0.900) & 1.110 (1.227) & 1.062 (1.175) & 1.190 (1.183) &  0.818 (0.818) \\
$3^+_1$ & 1.415 (1.303) & 1.152 (1.446) & 1.445 (1.269) & 1.255 (1.291) & 0.844 (-) \\
$4^+_2$ & 1.103 (1.395) & 1.304 (1.390) & 1.374 (1.384) & 1.406 (1.370) & 1.043 (-) \\
$5^+_1$ & 1.601 (1.508) & 1.340 (1.728) & 1.642 (1.533) & 1.446 (1.557) & 1.028 (-) \\
$0^+_2$ & 0.813 (0.828*) & 1.159 (1.174*) & 0.971 (1.199) & 1.098 (1.102) & 1.034 (1.034) \\
$2^+_3$ & 1.377 (1.227) & 1.243 (1.341) & 1.405 (1.496) & 1.220 (1.200) & 1.024 (-) \\
\hline
 \end{tabular}
\caption{
 Energies of the Hf isotopes. The theoretical values are listed for each nucleus 
in the first row of each column, while the experimental values are in the parenthesis. 
A star (*) indicates that the theory predicts a degeneracy of irreps. In this case, for the
experimental values an average, estimated values are given.
The experimental values, except for the $0^+$-states, are taken from \cite{brook}.
The information on the $0^+$-states are taken from \cite{ani2025}.
}
\vspace{0.2cm}
\label{Hf-energy}
\end{table}
\end{center}

\begin{center}
\begin{table}[h!]
\centering
\begin{tabular}{|c|c|c|c|c|c|}
\hline\hline
$J^P_i$ & $^{174}_{72}$Hf  & $^{176}_{72}$Hf & $^{178}_{72}$Hf & $^{180}_{72}$Hf
& $^{182}_{72}$Hf \\
\hline
\hline
$0^+_1$ & 0. (0.) & 0. (0.) & 0. (0.) & 0. (0.) & 0. (0.) \\
$0^+_2$ & 0.813 (0.828*) & 1.159 (1.174*) & 0.971 (1.199) & 1.098 (1.102) & 1.034 (1.034) \\
$0^+_3$ & 1.859 (-) & 1.163 (1.293) & 1.356 (1.668*) & 1.317 (1.316) & 1.265 (1.265*) \\
$0^+_4$ & 2.100 (-) & 1.659 (1.499) & 2.497 (-) & 2.726 (-) & 1.746 (-) \\
$0^+_5$ & 2.313 (-) & 2.303 (2,192)  & 2.933 (-) & 3.135 (-) & 4.549 (-) \\
\hline
 \end{tabular}
\caption{
$0^+$ spectrum of the Hf isotopes.
A star (*) indicates that the theory predicts a degeneracy of irreps. In this case, for the
experimental values an average, estimated values are given. The $0^+$ data are taken exclusively from \cite{ani2025}.
} 
\vspace{0.2cm}
\label{Hf-0}
\end{table}
\end{center}
\FloatBarrier

\section{$B(E2)$-values}
\label{appBE2}

\begin{center}
\begin{table}[h!]
\centering
\begin{tabular}{|c|c|c|c|c|c|}
\hline\hline
$J^P_i \rightarrow J^P_f$ & $^{146}_{62}$Sm  & $^{148}_{62}$Sm & $^{150}_{62}$Sm 
& $^{152}_{62}$Sm & $^{154}_{62}$Sm \\
\hline
\hline
$2^+_1$ $\rightarrow$ $0^+_1$ & 0.013 ($>$7.4) & 31. (31.) & 56.98 (57.) & 145. (145.) & 176. (176.)  \\
$2^+_2$ $\rightarrow$ $0^+_1$ & (2$>$3) 2.675 (3.30) & (2$>$3) 1.53 (-) & - (-) & - (-) \\
$2^+_2$ $\rightarrow$ $2^+_1$ & 0. (-) & (2$>$3) 2.45 (30.) & (2$>$3) 11.96 (-) & - (-) & - (-) \\
$2^+_2$ $\rightarrow$ $4^+_1$ & 0. (-) & (2$>$3) 0.161 (-) & (2$>$3) 49.43 (-) & - (-) & - (-)  \\
\hline
 \end{tabular}
\caption{
B(E2)-values for Sm isotopes. The cases, where the position of the second and first
$2^+$-states has to be interchanged, it is indicated in the table.
In each column the theoretical value is listed before the parenthesis and within the 
parenthesis the experimental value is given.
}
\vspace{0.2cm}
\label{Sm-BE2}
\end{table}
\end{center}

\begin{center}
\begin{table}[h!]
\centering
\begin{tabular}{|c|c|c|c|c|}
\hline\hline
$J^P_i$ $\rightarrow$ $J^P_f$ & $^{154}_{64}$Gd  & $^{156}_{64}$Gd & $^{158}_{64}$Gd 
& $^{160}_{64}$Gd \\
\hline
\hline
$2^+_1$ $\rightarrow$ $0^+_1$ & 157. (157.) & 189. (189.) & 198. (198.) & 200. (201) \\
$2^+_2$ $\rightarrow$ $0^+_1$ & 2.76 (2$>$3) (111.) & 9.81 (0.63) & 10.28 (3.4) & 9.25 (3.80) \\
$2^+_2$ $\rightarrow$ $2^+_1$ & 4.34 (2$>$3) (6.72) & 15.35 (3.3) & 16.08 (6.00) & 14.4 (1.07) \\
$2^+_2$ $\rightarrow$ $4^+_1$ & 0.27 (2$>$3) (19.4) & 1.0 (4.1) & 1.00 (0.27) & 0.88 (0.85) \\
\hline
 \end{tabular}
\caption{
B(E2)-values for Gd isotopes. The cases, where the position of the second and first
$2^+$-states has to be interchanged, it is indicated in the table.
In each column the theoretical value is listed before the parenthesis and within the 
parenthesis the experimental value is given.
}
\vspace{0.2cm}
\label{Gd-BE2}
\end{table}
\end{center}

\begin{center}
\begin{table}[h!]
\centering
\begin{tabular}{|c|c|c|c|}
\hline\hline
$J^P_i$ $\rightarrow$ $J^P_f$ & $^{160}_{66}$Dy  & $^{162}_{66}$Dy & $^{164}_{66}$Dy \\
\hline
\hline
$2^+_1$ $\rightarrow$ $0^+_1$ & 196. (196.) & 204. (202.) & 211.(211.) \\
$2^+_2$ $\rightarrow$ $0^+_1$ & 10.17 (4.46) & 9.44 (4.65) & 9.76 (4.30) \\
$2^+_2$ $\rightarrow$ $2^+_1$ & 15.91 (8.50) & 14.70 (8.23) & 15.20 (7.40)  \\
$2^+_2$ $\rightarrow$ $4^+_1$ & 0.98  (0.60) & 0.90 (0.63) & 0.93 (0.54) \\
\hline
 \end{tabular}
\caption{
B(E2)-values for Gd isotopes. In each column the theoretical value is listed before the parenthesis and within the 
parenthesis the experimental value is given.
}
\vspace{0.2cm}
\label{Dy-BE2}
\end{table}
\end{center}

\begin{center}
\begin{table}[h!]
\centering
\begin{tabular}{|c|c|c|c|c|c|}
\hline\hline
$J^P_i \rightarrow J^P_f$ & $^{162}_{68}$Er  & $^{164}_{68}$Er & $^{166}_{68}$Er & 
$^{168}_{68}$Er & $^{170}_{68}$Er \\
\hline
\hline
$2^+_1$ $\rightarrow$ $0^+_1$ & 189. (188.) & 206. (206.) & 217. (218.) & 213. (213.) 
& 207. (208.) \\
$2^+_2$ $\rightarrow$ $0^+_1$ &  9.81 (6.22) & 9.54 (5.3) & 10.05 (5.17) & 15.62 (4.68) 
& 0. (3.68) \\
$2^+_2$ $\rightarrow$ $2^+_1$ & 15.35 (14.70) & 14.84 (9.2) & 15.64 (9.6) & 24.40 ($$>$$8.) 
& 0. ($$>$$5.3) \\
$2^+_2$ $\rightarrow$ $4^+_1$ & 0.95 (1.80) & 0.91 (1.6) & 0.96 (0.78) & 1.50 (0.61) 
& 0.   
(0.29) \\
\hline
 \end{tabular}
\caption{
B(E2)-values for Er isotopes. In each column the theoretical value is listed before the parenthesis and within the 
parenthesis the experimental value is given.
}
\vspace{0.2cm}
\label{Er-BE2}
\end{table}
\end{center}

\begin{center}
\begin{table}[h!]
\centering
\begin{tabular}{|c|c|c|c|c|c|}
\hline\hline
$J^P_i \rightarrow J^P_f$ & $^{168}_{70}$Yb  & $^{170}_{70}$Yb & $^{172}_{70}$Yb 
& $^{174}_{70}$Yb & $^{176}_{70}$Yb \\
\hline
\hline
$2^+_1$ $\rightarrow$ $0^+_1$ & 209. (209.) & 201. (201.) & 212. (212.) & 201. (201.) 
& 183.(183) \\
$2^+_2$ $\rightarrow$ $0^+_1$ & 2.91 (5.0) & 0. (1.08) & 20.56 (2$>$3) (0.24) & 19.50 (-) 
& 12.03 (-) \\
$2^+_2$ $\rightarrow$ $2^+_1$ & 4.52 (9.2) & 0. (-) & 32.05 (2$>$3) (0.79) & 30.40 (-) 
& 18.69 (-) \\
$2^+_2$ $\rightarrow$ $4^+_1$ & 0.28 (1.8) & 0. (-) & 1.97 (2$>$3) (2.5) & 1.86 (-) 
& 1.14 (-) \\
\hline
 \end{tabular}
\caption{
B(E2)-values for Yb isotopes. In  $^{170}_{70}$Yb the lowest irrep is (36,0) and,
therefore, the $K=2$ band is not  present in this lowest irrep, i.e., no transitions.In each column the theoretical value is listed before the parenthesis and within the 
parenthesis the experimental value is given.
}
\vspace{0.2cm}
\label{Yb_BE2}
\end{table}
\end{center}

\begin{center}
\begin{table}[h!]
\centering
\begin{tabular}{|c|c|c|c|c|c|}
\hline\hline
$J^P_i \rightarrow J^P_f$ & $^{174}_{72}$Hf  & $^{176}_{72}$Hf & $^{178}_{72}$Hf 
& $^{180}_{72}$Hf & $^{182}_{72}$Hf \\
\hline
\hline
$2^+_1$ $\rightarrow$ $0^+_1$ & 152. (152.) & 184. (183.) & 160. (160.) & 154. (154.) 
& 174. (-) \\
$2^+_2$ $\rightarrow$ $0^+_1$ & 15.52 (2$>$3) (2.1) & 8.52 (0.98) & (2$>3$) 4.21 (3.9) 
& 4.05 (2$>$3) (-) & 36.51 (-) \\
$2^+_2$ $\rightarrow$ $2^+_1$ & 25.90 (2$>$3) (-) & 13.26 (-) & 6.28 (2$>$3) (4.4) & 6.28 (-) 
&  57.80 (-) \\
$2^+_2$ $\rightarrow$ $4^+_1$ & 1.61 (2$>$3) (13.) & 0.81 (5.7) & (2$>$3) 0.40 (0.26) 
& 0.38 (2$>$3)  (-)  & 3.67 (-)  
 \\
\hline
 \end{tabular}
\caption{
B(E2)-values for Hf isotopes. Due to the near degeneracy of the $2_{2,3}^+$ states, in 
$^{174}_{72}$Hf these states have to be interchanged in character. The same change
of characteristic has been applied to $^{178}_{72}$Hf and$^{180}_{72}$Hf, 
because the $2_3^+$ state 
connects to the ground state, but the $2_2^+$ state does not.
In each column the theoretical value is listed before the parenthesis and within the 
parenthesis the experimental value is given.
}
\vspace{0.2cm}
\label{Hf-BE2}
\end{table}
\end{center}

\end{appendices}

\end{document}